\documentclass[11pt]{article}
\usepackage{amssymb,amsmath,amsfonts}
\usepackage{graphicx}
\usepackage{graphics}
\usepackage{eepic,epsfig}

\makeatletter
\@addtoreset{equation}{section}

\makeatother

\begin{document}

\title{Electromagnetic Wightman functions and vacuum densities\\
for a brane intersecting the AdS boundary}
\author{A. A. Saharian\thanks{%
E-mail: saharian@ysu.am}\, , R. M. Avagyan, V. F. Manukyan \vspace{0.3cm} \\
%EndAName
\textit{Institute of Physics, Yerevan State University,}\\
\textit{1 Alex Manoogian Street, 0025 Yerevan, Armenia}}
\maketitle

\begin{abstract}
We investigate the combined effects of a brane intersecting the AdS boundary
and background gravitational field on the local characteristics of the
electromagnetic vacuum. Two types of boundary conditions on the brane are
considered, which are higher-dimensional generalizations of the perfect
electric (PEC) and perfect magnetic (PMC) boundary conditions in Maxwell's
electrodynamics. The brane-induced contributions to the Wightman functions
of the vector potential and field tensor are explicitly extracted. Simple
expressions in terms of elementary functions are provided. The behavior of
the vacuum expectation values (VEVs) is mimicked by a scalar field with a
negative effective mass squared determined by the radius of the AdS
spacetime. The expectation values of the electric and magnetic fields
squares and of the energy-momentum tensor are investigated as local
characteristics of the vacuum state. The brane-induced contributions to
these VEVs have opposite signs for the PEC and PMC conditions. For the PMC
condition, this contribution is negative for the electric field squared and
positive for the magnetic field squared. The VEV of the energy-momentum
tensor has a nonzero off-diagonal component. The brane-induced vacuum energy
density is positive for PMC condition, whereas the normal and parallel
stresses change sign as functions of the distance from the brane. Unlike the
problem involving a planar boundary in the Minkowski bulk, the vacuum
energy-momentum tensor does not vanish in (3+1)-dimensional AdS spacetime.
\end{abstract}

\vspace{0.3cm}


\section{Introduction}

In the absence of a theory of quantum gravity, the current investigations of
the influence of gravitational fields on quantum phenomena are conducted
within the framework of semiclassical theory, where the gravitational field
is considered as a classical spacetime background. The characteristic
energies of quantum phenomena in gravitational fields are expected to be
much larger than those for the fields in the Standard Model, and this
approach has a fairly wide range of applications. Interesting results from
research in this area include the polarization of the vacuum of quantum
fields and the particle creation in strong gravitational fields. These
phenomena play an important role in the cosmology of the early universe and
in the physics of black holes. Notably, the vacuum expectation values (VEVs)
of the energy-momentum tensor formed as a result of vacuum polarization
generally do not satisfy the energy conditions of the singularity theorems,
and this may be the key to solving the singularity problem in classical
gravitational theories.

In general, the presence of gravitational fields reduces the symmetry of
problems involving quantum fields, and exact results for physical
characteristics are obtained only for highly symmetric background
geometries. The dynamics of quantum fields in de Sitter (dS) and anti-de
Sitter (AdS) spacetime background geometries, in particular, is a subject of
active research (see \cite{Hawk73,Grif09} for geometrical properties). Along
with Minkowski space, these manifolds are maximally symmetric, and the
results obtained for them provide qualitative insight into the influence of
gravitational fields on quantum effects in more complicated geometries. In
the present paper, for AdS background geometry, we investigate combined
effects of boundaries and gravitational field on local characteristics of
the electromagnetic vacuum in general number of spatial dimensions $D$. In
addition to high symmetry, our choice of AdS spacetime has several other
motivations. The early interest in this geometry was driven by fundamental
questions related to the quantization of fields in gravitational
backgrounds. These investigations became more important when it has been
shown that AdS spacetime emerges as a ground state in supergravity and
string theories and as the near-horizon geometry for extremal black holes.
The current growing interest in problems with the AdS background geometry is
also due to the fact that it appears as a key element in two promising
developments of modern theoretical physics, namely, in AdS/CFT
correspondence and braneworld models with large extra dimensions.

The AdS/CFT correspondence establishes holographic duality between
supergravity and string theories on the AdS bulk and the conformal field
theory (CFT) localized on its boundary (for reviews see \cite{Ahar00}-\cite%
{Zaa15}). It provides a possibility to study phenomena in one theory based
on the features of the dynamics in the dual theory and serves as a powerful
tool for investigating effects in the strong coupling regime of gauge
theories. This has found interesting applications in both high-energy and
condensed matter physics. The braneworld models \cite{Brax04,Maar10}
naturally appear in the string theory context and provide a geometric
explanation of the hierarchy problem between electroweak and gravitational
energy scales. They open new perspectives for addressing problems in
particle physics, gravity, and cosmology. Braneworlds present an example of
a physical theory in which the dynamics of a part of the degrees of freedom
is localized on a hypersurface (brane) in a high-dimensional spacetime. In
models on the AdS bulk the weakness of the effective gravity on the visible
brane is closely related to the negative curvature of the AdS spacetime.
Note that this negative curvature also serves as a natural infrared
regulator for quantum fields propagating on AdS background.

A common feature in both setups of AdS/CFT correspondence and braneworlds is
the presence of boundaries. Their interaction with bulk quantum fields
produces interesting effects that significantly impact the dynamics of the
fields. In particular, boundary conditions imposed on fields lead to a
change in the spectrum of zero-point fluctuations. This results in
contributions to the VEVs of physical quantities that depend on the location
and geometry of the boundaries. This is a manifestations of the Casimir
effect that has been studied in detail in the literature for different types
of bulk and boundary geometries and fields \cite{Most97}-\cite{Casi11}. The
interest to the investigations of the Casimir effect in branworlds was
motivated by the need to stabilize the location of the branes, as well as by
the possibility of generating a cosmological constant on branes. The Casimir
energy and the respective effective potential for the radion field have been
studied within the framework of Randall-Sundrum braneworld models with
planar branes and in their higher dimensional generalizations with compact
subspaces \cite{Fabi00}-\cite{Teo13}. More detailed information on the
vacuum fluctuations of bulk fields is contained in the expectation values of
the energy momentum tensor. The corresponding VEVs for scalar and fermionic
fields were discussed in \cite{Knap04}-\cite{Bell22AdS} for the bulk
energy-momentum tensor and in \cite{Saha04,Saha06c} for the surface
energy-momentum tensor localized on the branes. The investigations for
global and local characteristics of the electromagnetic vacuum in
braneworlds on AdS bulk are presented in \cite{Teo10}-\cite{Saha20}.

In the references cited above, the investigations were performed for the
geometry of branes parallel to the AdS boundary. In an extension of the
AdS/CFT correspondence, called AdS/BCFT correspondence, the conformal field
theory contains boundaries \cite{Taka11,Fuji11,Karc01} (see also \cite%
{Suzu24}-\cite{Hao25} and references therein for recent discussions and
developments). In those problems the boundaries/branes are present in the
dual theory on the AdS bulk which intersect the AdS boundary
(end-of-the-world (EOW) brane). The AdS/BCFT duality provides a consistent
framework for studying the effects of interfaces, impurities and topological
defects in quantum field theories and has found interesting applications in
condensed matter physics. The surfaces crossing the AdS boundary are an
inherent feature of the procedure for evaluating the entanglement entropy of
bounded quantum systems in conformal field theories suggested in \cite%
{Ryu06,Ryu06b} (for reviews see \cite{Nish09,Chen22}). In this scheme, the
entanglement entropy is expressed in terms of the area of the minimal
surface in the AdS bulk anchored at the boundary of the dual conformal field
theory. The boundaries in the AdS bulk are sources of Casimir-type
contributions in the expectation values of physical observables for bulk
fields. In \cite{Beze15,Bell22,Saha23}, the bulk and surface Casimir
densities for a scalar field were studied on the AdS bulk in the geometry of
planar branes orthogonal to the AdS boundary. Continuing in this line of
investigations, in the present paper we consider an exactly solvable problem
for the polarization of the electromagnetic vacuum in the presence of a
planar brane intersecting the AdS boundary.

The organization of the paper is as follows. The next section presents the
problem setup and the complete set of electromagnetic modes used in the
canonical quantization procedure. In Sections \ref{sec:WF} and \ref{sec:FT},
these modes are used for evaluating the two-point functions of the vector
potential and field tensor. The brane-induced contributions are explicitly
separated. The VEVs of the electric and magnetic fields squares and photon
condensate are studied in Section \ref{sec:E2B2}. Section \ref{sec:EMT} is
devoted to the investigation of the VEVs for the energy-momentum tensor. The
main results are summarized in Section \ref{sec:Conc}. The evaluation of the
integral in the expressions of the two-point functions is presented in
Section \ref{sec:Int}.

\section{Setup and the electromagnetic modes}

\label{sec:Setup}

As the background geometry we consider $(D+1)$ -dimensional AdS spacetime
generated by a negative cosmological constant $\Lambda $. In the Poincar\'{e}
coordinates $(x^{0}=t,x^{1},...,x^{D-1},x^{D}=z)$, the line element reads 
\begin{equation}
ds^{2}=g_{\mu \nu }dx^{\mu }dx^{\nu }=(\alpha /z)^{2}\eta _{\mu \nu }dx^{\mu
}dx^{\nu },  \label{ds2}
\end{equation}%
where $\eta _{\mu \nu }=\mathrm{diag}(1,-1,...-1)$\ is the Minkowskian
metric tensor. The AdS radius $\alpha $ is given by $\alpha
^{2}=-D(D-1)/(2\Lambda )$. In (\ref{ds2}), $0\leq z<\infty $ and the
hypersurfaces $z=0$ and $z=\infty $ correspond to the boundary and horizon
of the AdS spacetime. The proper distance along the axis $z$ is measured by
the coordinate $y$ that is connected to the $z$-coordinate by the relation $%
y=\alpha \ln (z/\alpha )$. We consider a quantum electromagnetic field with
the vector potential $A_{\mu }(x)$ and the field tensor $F_{\mu \nu
}=\partial _{\mu }A_{\nu }-\partial _{\nu }A_{\mu }$. In the absence of
external sources we have the field equation%
\begin{equation}
\nabla _{\nu }F^{\mu \nu }=\frac{1}{\sqrt{|g|}}\partial _{\nu }\left( \sqrt{%
|g|}F^{\mu \nu }\right) =0,  \label{Feq}
\end{equation}%
with $g$ being the determinant of the metric tensor $g_{\mu \nu }$. The
vector potential will be fixed by the gauge conditions $A_{D}=0$ (radial
gauge) and $\nabla _{\mu }A^{\mu }=0$. By taking into account that the
metric tensor only depends on the coordinate $x^{D}$, these to conditions
are reduced to the gauge condition $\partial _{\mu }A^{\mu }=0$. Note that
the imposed constraints do not uniquely fix the vector potential. An
additional gauge transformation $A_{\mu }^{\prime }=A_{\mu }+\partial _{\mu
}\chi $ can be made with a function $\chi =\chi (x)$ obeying the conditions $%
\partial _{D}\chi (x)=0$ and $g^{\mu \nu }\partial _{\mu }\partial _{\nu
}\chi (x)=0$.

We are interested in how a codimension-one boundary at $x^{1}=0$ affects the
properties of the Poincar\'{e} vacuum for the electromagnetic field. The
boundary's physical nature can differ. Here, considering possible
applications in braneworld models and AdS/BCFT correspondence, we will
simply refer to the boundary as a "brane". Two types of boundary conditions
will be considered for the field on the boundary. These are
higher-dimensional generalizations of perfect magnetic conductor (PMC) and
perfect electric conductor (PEC) conditions in 3D Maxwell's theory. In terms
of the field tensor, the boundary conditions are written as 
\begin{equation}
\left. n^{\mu }F_{\mu \nu }\right\vert _{x^{1}=0}=0,  \label{PM}
\end{equation}%
and 
\begin{equation}
\left. n^{\mu _{1}\ast }F_{\mu _{1}...\mu _{D-1}}\right\vert _{x^{1}=0}=0,
\label{PE}
\end{equation}%
for the PMC and PEC conditions, respectively. Here, $n^{\mu }$ is the normal
to the brane and $^{\ast }F_{\mu _{1}...\mu _{D-1}}=\varepsilon _{\mu \nu
\mu _{1}...\mu _{D-1}}F^{\mu \nu }/(D-1)!$ is the dual of the field tensor.
For the geometry under consideration we have $n^{\mu }=\pm \delta _{1}^{\mu
}z/\alpha $, where the upper and lower signs correspond to the regions $%
x^{1}\geq 0$ and $x^{1}\leq 0$, respectively. Both the boundary conditions (%
\ref{PM}) and (\ref{PE}) ensure ideal reflection for all polarization modes.
Note that this is not the case for the boundary condition (\ref{PE}) imposed
on a massive vector field (Proca field; see, for example, \cite%
{Davi81,Bart85,Saha26}). The PEC condition does not restrict longitudinally
polarized modes, and these modes do not experience the presence of the
boundary.

The information of the local properties of the vacuum state is encoded in
two-point functions. As such, we will consider the positive frequency
Wightman function. It can be evaluated by the summation over complete set of
electromagnetic modes obeying the boundary conditions. In accordance of the
problem symmetry, the electromagnetic modes for the vector potential can be
presented in the factorized form $A_{(\beta )\mu }=f_{\mu
}(z)(C_{1}e^{ik^{1}x^{1}}+C_{2}e^{-ik^{1}x^{1}})e^{i(\mathbf{k}_{\Vert
}\cdot \mathbf{x}_{\Vert }-\omega t)}$, where $\beta $ stands for the set of
quantum numbers specifying the modes, $\mathbf{k}_{\Vert }=(k^{2},\ldots
,k^{D-1})$, $\mathbf{x}_{\Vert }=(x^{2},\ldots ,x^{D-1})$, and $\mathbf{k}%
_{\Vert }\cdot \mathbf{x}_{\Vert }=\sum_{i=2}^{D-1}k^{i}x^{i}$. From the
Maxwell's equations we can see that for normalizable modes $f_{\mu
}(z)\varpropto $ $z^{D/2-1}J_{D/2-1}(\lambda z)$, where $J_{\nu }(x)$ is the
Bessel function, $\lambda =\sqrt{\omega ^{2}-k^{2}}$ and $%
k^{2}=\sum_{i=1}^{D-1}(k^{i})^{2}$. The ratio of the coefficients $C_{1}$
and $C_{2}$ is determined from the boundary conditions at $x^{1}=0$. The
mode functions become%
\begin{equation}
A_{(\beta )\mu }(x)=i^{\delta _{1\mu }}C_{\beta }e_{(\sigma )\mu
}z^{D/2-1}J_{D/2-1}(\lambda z)h\left( k^{1}x^{1}-\pi \delta _{1\mu
}/2\right) e^{i(\mathbf{k}_{\Vert }\cdot \mathbf{x}_{\Vert }-\omega t)},
\label{Amu}
\end{equation}%
where $\mu =0,1,2,\ldots ,D$, and 
\begin{equation}
h(u)=\left\{ 
\begin{array}{cc}
\cos (u), & \text{PMC condition} \\ 
\sin (u), & \text{PEC condition}%
\end{array}%
\right. .  \label{hu}
\end{equation}%
We have $D-1$ transverse polarization states enumerated by $\sigma =1,\ldots
,D-1$, with the polarization vectors $e_{(\sigma )\mu }$ constrained by the
conditions%
\begin{equation}
\eta ^{\mu \rho }e_{(\sigma )\mu }e_{(\sigma ^{\prime })\rho }=-\delta
_{\sigma \sigma ^{\prime }},\;k^{\mu }e_{(\sigma )\mu }=0,\;e_{(\sigma )D}=0.
\label{epol}
\end{equation}%
Below, in the evaluation of the two-point functions, for the summation over
polarizations we will use the relation 
\begin{equation}
\sum_{\sigma =1}^{D-1}e_{(\sigma )\mu }e_{(\sigma )\nu }=\frac{k_{\mu
}k_{\nu }}{\lambda ^{2}}-\eta _{\mu \nu },  \label{sumpol}
\end{equation}%
for the polarization vector. The complete set of quantum numbers specifying
the modes is given by $\beta =\left( \sigma ,\lambda ,k^{1},\mathbf{k}%
_{\Vert }\right) $ with $0\leq \lambda <\infty $ and $0\leq k^{1}<\infty $.

The constants $C_{\beta }$ in (\ref{Amu}) are determined from the
normalization condition 
\begin{equation}
\int d^{D}x\,\left( \alpha /z\right) ^{D-1}A_{(\beta ^{\prime })}^{\ast \mu
}A_{(\beta )\mu }=-\frac{2\pi }{\omega }\delta _{\beta \beta ^{\prime }}.
\label{NC}
\end{equation}%
Substituting the mode functions and using the normalization condition for
the polarization vectors we get%
\begin{equation}
\left\vert C_{\beta }\right\vert ^{2}=\frac{4\lambda }{\left( 2\pi \right)
^{D-2}\alpha ^{D-3}\omega }.  \label{C2}
\end{equation}%
In spatial dimension $D=3$ the electromagnetic field is conformally
invariant. In this case we have $z^{D/2-1}J_{D/2-1}(\lambda z)=\sqrt{2/\pi
\lambda }\sin \left( \lambda z\right) $ and the modes (\ref{Amu}) reduce to
the corresponding mode functions on the Minkowski spacetime with the line
element $ds_{\mathrm{M}}^{2}=\eta _{\mu \nu }dx^{\mu }dx^{\nu }$. In the
problem on the Minkowski bulk there are two boundaries. The first one
corresponds to the planar boundary $x^{1}=0$ with PMC or PEC conditions and
the second one, located at $z=0$, is the conformal image of the AdS boundary
with the PEC boundary condition on the electromagnetic field.

As mentioned above, there is a possibility of additional gauge
transformation that maintains the conditions $A_{D}=0$ and $\partial _{\mu
}A^{\mu }=0$. The corresponding transformation function $\chi (x)$ does not
depend on the coordinate $z$. Taking it in the form $\chi (x)=\chi
_{1}\left( x^{1}\right) e^{i(\mathbf{k}_{\Vert }\cdot \mathbf{x}_{\Vert
}-\omega t)}$, from the constraint $g^{\mu \nu }\partial _{\mu }\partial
_{\nu }\chi (x)=0$ we get $\chi _{1}\left( x^{1}\right) =\sum_{j=\pm
}c_{j}e^{ji\tilde{k}^{1}x^{1}}$ with constants $c_{\pm }$ and $\tilde{k}^{1}=%
\sqrt{\omega ^{2}-\mathbf{k}_{\Vert }^{2}}$. The gauge transformed vector
potential becomes $A_{\mu }^{\prime }=A_{\mu }+\partial _{\mu }\chi $ where $%
\partial _{\mu }\chi =i\tilde{k}^{\mu }e^{i(\mathbf{k}_{\Vert }\cdot \mathbf{%
x}_{\Vert }-\omega t)}\sum_{j=\pm }j^{\delta _{1\mu }}c_{j}e^{ji\tilde{k}%
^{1}x^{1}}$ and $\tilde{k}^{l}=k^{l}$ for $l=2,3,\ldots ,D-1$. However, for
the modes $A_{\mu }=A_{(\beta )\mu }(x)$ from (\ref{Amu}), the mode
functions $A_{\mu }^{\prime }=A_{(\beta )\mu }^{\prime }(x)$ are not
normalizable. Therefore, the gauge conditions $A_{D}=0$ and $\partial _{\mu
}A^{\mu }=0$, combined with the requirement of normalizability, uniquely fix
the mode functions (\ref{Amu}).

In the mode functions (\ref{Amu}), we took the Bessel function as a solution
to the Bessel equation. This corresponds to the standard Dirichlet
quantization. For $D>3$, the normalizability condition uniquely determines
the modes (\ref{Amu}) as a complete set used in the canonical quantization
procedure. In the special cases $D=2,3$, the modes with the linear
combination $C_{\beta }J_{D/2-1}(\lambda z)+B_{\beta }Y_{D/2-1}(\lambda z)$
are also normalizable. Here, $C_{\beta }$ and $B_{\beta }$ are constants,
and $Y_{D/2-1}(\lambda z)$ is the Neumann function. Different choices of the
ratio $B_{\beta }/C_{\beta }$ correspond to different quantization
procedures and different theories on the AdS boundary in AdS/CFT
correspondence. The special case $C_{\beta }=0$ corresponds to the Neumann
quantization. The more general case with $C_{\beta },B_{\beta }\neq 0$
realizes a mixed (Robin) condition on the AdS boundary.

The mode functions for the field tensor read%
\begin{align}
F_{(\beta )\mu \nu }(x)=& i^{\delta _{1\mu }}C_{\beta }z^{\frac{D}{2}-1}J_{%
\frac{D}{2}-1-\delta _{\nu D}}(\lambda z)h\left( k^{1}x^{1}-\pi \delta
_{1\mu }/2\right) e^{i(\mathbf{k}_{\Vert }\cdot \mathbf{x}_{\Vert }-\omega
t)}  \notag \\
& \times \left\{ 
\begin{array}{ll}
i\left[ \eta _{\nu \rho }e_{(\sigma )\mu }-\eta _{\mu \rho }e_{(\sigma )\nu }%
\right] k^{\rho }, & \nu \neq D \\ 
-e_{(\sigma )\mu }\lambda , & \nu =D%
\end{array}%
\right. ,  \label{Fmodes}
\end{align}%
where $\mu \neq D$. For $D>3$, we have $F_{(\beta )\mu \nu }(x)=0$ at $z=0$
and, hence, both the PMC and PEC boundary conditions are obeyed on the AdS
boundary. In spatial dimension $D=3$ one has $F_{(\beta )\mu \nu
}(x)|_{z=0}=0$ for $\mu ,\nu \neq D$, and $F_{(\beta )D\nu }(x)|_{z=0}\neq 0$
for $\nu \neq D$. For this case, on the AdS boundary the modes obey the PEC
condition and the PMC condition is not obeyed. For $D=2$ we have the
opposite situation: the modes obey the PMC condition, $F_{(\beta )2\nu
}(x)|_{z=0}=0$ for $\nu \neq 2$, and the PEC condition is not satisfied, $%
F_{(\beta )\mu \nu }(x)|_{z=0}\neq 0$ for $\mu ,\nu \neq 2$.

Note that the choice of the Bessel function $J_{\nu }(x)$ in (\ref{Amu}) is
dictated by the normalizability condition of the modes used in the canonical
quantization procedure. There are also non-normalizable modes. The radial
part of these modes is a linear combination of the Bessel and Neumann
functions and the relative coefficient in this combination is determined by
a boundary condition at $z=0$. Although the irregular modes are not included
in the quantization scheme, they play an important role in the AdS/CFT
correspondence. These modes serve as sources of currents in the conformal
field theory localized on the AdS boundary.

\section{Wightman function of the vector potential}

\label{sec:WF}

The positive-frequency Wightman function (Wightman bitensor) of the vector
potential is defined as the expectation value $\langle 0|A_{\mu }(x)A_{\nu
}(x^{\prime })|0\rangle =\left\langle A_{\mu }A_{\nu ^{\prime
}}\right\rangle $ for the vacuum state $|0\rangle $. Expanding the operator $%
A_{\mu }(x)$ over the complete set of the mode functions (\ref{Amu}) and by
taking into account that the vacuum state is nullified by the annihilation
operator, the mode-sum formula 
\begin{equation}
\left\langle A_{\mu }A_{\nu ^{\prime }}\right\rangle =\sum_{\sigma
=1}^{D-1}\int d\mathbf{k}_{\Vert }\int_{0}^{\infty }d\lambda
\,\int_{0}^{\infty }dk^{1}A_{(\beta )\mu }\left( x\right) A_{(\beta )\nu
}^{\ast }(x^{\prime }),  \label{AA}
\end{equation}%
is obtained. Substituting the mode functions, summing over the
polarizations, and by using the relation 
\begin{equation}
h(u)h(u^{\prime })=\frac{1}{2}\left[ \cos (u-u^{\prime })+\delta _{\mathrm{b}%
}\cos (u+u^{\prime })\right] ,  \label{hh}
\end{equation}%
with%
\begin{equation}
\delta _{\mathrm{b}}=\left\{ 
\begin{array}{cc}
1, & \text{PMC condition} \\ 
-1, & \text{PEC condition}%
\end{array}%
\right. ,  \label{delb}
\end{equation}%
the function (\ref{AA}) is decomposed into two contributions. The first one
comes from $\cos (u-u^{\prime })$ and corresponds to the Wightman function
in the problem where the brane is absent (brane-free problem, for the gauge
boson propagator in AdS spacetime see \cite{Alle86,Liu99,Hoke99}). We denote
this part by $\left\langle A_{\mu }A_{\nu ^{\prime }}\right\rangle _{0}$.
The second contribution comes from $\cos (u+u^{\prime })$ and is the
brane-induced contribution. Denoting it by $\left\langle A_{\mu }A_{\nu
^{\prime }}\right\rangle _{\mathrm{b}}$, we get 
\begin{equation}
\left\langle A_{\mu }A_{\nu ^{\prime }}\right\rangle =\left\langle A_{\mu
}A_{\nu ^{\prime }}\right\rangle _{0}+\left\langle A_{\mu }A_{\nu ^{\prime
}}\right\rangle _{\mathrm{b}},  \label{AAdec}
\end{equation}%
with the representation 
\begin{align}
\left\langle A_{\mu }A_{\nu ^{\prime }}\right\rangle _{\mathrm{b}}& =\frac{%
2\delta _{\mathrm{b}}\left( zz^{\prime }\right) ^{\frac{D}{2}-1}}{\left(
2\pi \right) ^{D-2}\alpha ^{D-3}}\int_{0}^{\infty }d\lambda \,\lambda J_{%
\frac{D}{2}-1}(\lambda z)J_{\frac{D}{2}-1}(\lambda z^{\prime })\int d\mathbf{%
k}_{\Vert }\int_{0}^{\infty }dk^{1}\frac{1}{\omega }  \notag \\
& \times \left( \frac{k_{\mu }k_{\nu ^{\prime }}}{\lambda ^{2}}-\eta _{\mu
\nu ^{\prime }}\right) e^{i(\mathbf{k}_{\Vert }\cdot \Delta \mathbf{x}%
_{\Vert }-\omega \Delta t)}\cos \left[ k^{1}x_{+}^{1}-\pi \left( \delta
_{1\mu }+\delta _{1\nu }\right) /2\right] ,  \label{AA2}
\end{align}%
for $\mu ,\nu \neq D$. Here, $\omega =\sqrt{\lambda ^{2}+k^{2}}$ and $%
x_{+}^{1}=x^{1}+x^{\prime 1}$. The components $\left\langle A_{\mu }A_{\nu
^{\prime }}\right\rangle _{\mathrm{b}}$ with one of the indices being equal
to $D$ become zero. The problem is symmetric with respect to the plane $%
x^{1}=0$ and in the discussion below the results will be given for the
region $x^{1}>0$.

For further transformation, we write the two-point function in the form%
\begin{equation}
\left\langle A_{\mu }A_{\nu ^{\prime }}\right\rangle _{\mathrm{b}}=\frac{%
2\delta _{\mathrm{b}}\left( zz^{\prime }\right) ^{\frac{D}{2}-1}}{\left(
2\pi \right) ^{D-2}\alpha ^{D-3}}\int_{0}^{\infty }d\lambda \,\lambda J_{%
\frac{D}{2}-1}(\lambda z)J_{\frac{D}{2}-1}(\lambda z^{\prime })\left( \frac{%
\partial _{\mu }\partial _{\nu }^{\prime }}{\lambda ^{2}}-c_{1}\eta _{\mu
\nu }\right) I(x,x^{\prime }),  \label{AA3}
\end{equation}%
where $\partial _{\nu }^{\prime }=\partial /\partial x^{\prime \nu }$, $%
c_{1}=-1$ for $\mu ,\nu =1$ and $c_{1}=1$ in other cases. Here we have
introduced the notation%
\begin{equation}
I(x,x^{\prime })=\int d\mathbf{k}_{\Vert }e^{i\mathbf{k}_{\Vert }\cdot
\Delta \mathbf{x}_{\Vert }}\int_{0}^{\infty }dk^{1}\frac{e^{-i\omega \Delta
t}}{\omega }\cos \left( k^{1}x_{+}^{1}\right) ,  \label{Ixx}
\end{equation}%
for the integral over the momentum. The evaluation of this integral is
presented in Appendix \ref{sec:Int} with the result given by (\ref{Ixx4}).
For the components with $\mu ,\nu \neq D$ this gives%
\begin{equation}
\left\langle A_{\mu }A_{\nu ^{\prime }}\right\rangle _{\mathrm{b}}=\frac{%
2\delta _{\mathrm{b}}\alpha ^{3-D}}{\left( 2\pi \right) ^{\frac{D}{2}-1}}%
\left[ \partial _{\mu }\partial _{\nu }^{\prime }\mathcal{J}_{\frac{D}{2}%
-1}^{(-1)}(z,z^{\prime },s_{+})-c_{1}\eta _{\mu \nu ^{\prime }}\mathcal{J}_{%
\frac{D}{2}-1}^{(1)}(z,z^{\prime },s_{+})\right] ,  \label{AA4}
\end{equation}%
where 
\begin{equation}
s_{+}=\sqrt{(x_{+}^{1})^{2}+|\Delta \mathbf{x}_{\Vert }|^{2}-\left( \Delta
t\right) ^{2}},  \label{spl}
\end{equation}%
and%
\begin{equation}
\mathcal{J}_{b}^{(n)}(z,z^{\prime },s_{+})=\int_{0}^{\infty }d\lambda
\,\lambda ^{n}j_{b}(\lambda z)j_{b}(\lambda z^{\prime })f_{b}(\lambda s_{+}).
\label{Jcal}
\end{equation}%
The remaining components with $\mu =D$ or $\nu =D$ become zero. Here, we
have introduced the notations%
\begin{equation}
j_{b}(x)=x^{b}J_{b}(x),\;f_{b}(x)=K_{b}(x)/x^{b},  \label{fnuK}
\end{equation}%
with $K_{b}(x)$ being the Macdonald function.

The integral (\ref{Jcal}) for $n=1$ is evaluated by using the formula from 
\cite{Prud2}:%
\begin{equation}
\mathcal{J}_{b}^{(1)}(z,z^{\prime },s_{+})=\frac{2^{b-1}\Gamma \left( b+%
\frac{1}{2}\right) }{\sqrt{\pi }zz^{\prime }\left( u_{+}^{2}-1\right) ^{b+%
\frac{1}{2}}},\;u_{+}=\frac{z^{2}+z^{\prime 2}+s_{+}^{2}}{2zz^{\prime }}.
\label{Jcal1}
\end{equation}%
The derivatives appearing in (\ref{AA4}) are evaluated using the relations 
\begin{equation}
j_{b}^{\prime }(x)=xj_{b-1}(x),\;f_{b}^{\prime }(x)=-xf_{b+1}(x).
\label{jfder}
\end{equation}%
Additionally, using the representation (\ref{Jcal1}) shows that%
\begin{equation}
\partial _{s_{+}}\mathcal{J}_{b}^{(1)}(z,z^{\prime },s_{+})=-\frac{s_{+}u_{+}%
}{zz^{\prime }}\mathcal{J}_{b+1}^{(1)}(z,z^{\prime },s_{+}).  \label{Jcalder}
\end{equation}%
In the AdS spacetime, the invariant (chordal) distance $Z(x,x^{\prime })$
between the spacelike separated points $x$ and $x^{\prime }$ is expressed in
terms of the geodesic distance $\sigma (x,x^{\prime })$ as $Z(x,x^{\prime
})=\cosh \!\left( \sigma (x,x^{\prime })/\alpha \right) -1$. The geodesic
distance is given by $\sigma (x,x^{\prime })=-\eta _{\mu \nu }\Delta x^{\mu
}\Delta x^{\nu }/(2zz^{\prime })$, where $\Delta x^{D}=z-z^{\prime }$. Now
we see that the relation $u_{+}=1+\sigma (x,x_{\mathrm{im}}^{\prime })$
takes place, where $x_{\mathrm{im}}^{\prime }=(t,-x^{\prime 1},x^{\prime
2},\ldots ,x^{\prime D-1})$ is the image of the point $x^{\prime }$ with
respect to the plane $x^{1}=0$.

An important object in the AdS/BCFT correspondence is the bulk-to-boundary
propagator. The evaluation of this two-point function is usually performed
in Euclidean signature. The corresponding procedure for the electromagnetic
field is similar to that discussed in the literature for the case of a
scalar field (see, e.g., \cite{Alis11,Hint15} for Dirichlet and Neumann
boundary conditions and \cite{Beze15} in the case of Robin condition). This
procedure is based on the Euclidean analog of the modes (\ref{Amu}).

\section{Wightman function for the field tensor}

\label{sec:FT}

Having the Wightman function of the vector potential we can evaluate the
corresponding two-point function for the field tensor, $\langle 0|F_{\sigma
\mu }(x)F_{\rho \nu }(x^{\prime })|0\rangle =\left\langle F_{\sigma \mu
}F_{\rho ^{\prime }\nu ^{\prime }}\right\rangle $. It is decomposed into
brane-free and brane-induced contributions as 
\begin{equation}
\left\langle F_{\sigma \mu }F_{\rho ^{\prime }\nu ^{\prime }}\right\rangle
=\left\langle F_{\sigma \mu }F_{\rho ^{\prime }\nu ^{\prime }}\right\rangle
_{0}+\left\langle F_{\sigma \mu }F_{\rho ^{\prime }\nu ^{\prime
}}\right\rangle _{\mathrm{b}}.  \label{FFdec}
\end{equation}%
From (\ref{AA4}), for the components with $\sigma ,\mu ,\rho ,\nu \neq 1,D$
one finds 
\begin{equation}
\left\langle F_{\sigma \mu }F_{\rho ^{\prime }\nu ^{\prime }}\right\rangle _{%
\mathrm{b}}=\frac{8\delta _{\mathrm{b}}\alpha ^{3-D}}{\left( 2\pi \right) ^{%
\frac{D}{2}-1}}\left[ \partial _{\lbrack \sigma }\partial _{\mu ]}\partial
_{\lbrack \rho ^{\prime }}\partial _{\nu ^{\prime }]}\mathcal{J}_{\frac{D}{2}%
-1}^{(-1)}(z,z^{\prime },s_{+})-\eta _{\lbrack \mu \lbrack \nu ^{\prime
}}\partial _{\sigma ]}\partial _{\rho ^{\prime }]}\mathcal{J}_{\frac{D}{2}%
-1}^{(1)}(z,z^{\prime },s_{+})\right] .  \label{FFb}
\end{equation}%
where the antisymmetrization is done between the primed and unprimed pairs
of indices in accordance with%
\begin{equation}
a_{[\sigma }b_{\mu ]}=\frac{1}{2}\left( a_{\sigma }b_{\mu }-a_{\mu
}b_{\sigma }\right) .  \label{antsym}
\end{equation}%
Now, by taking into account that the partial derivatives commute, we
conclude that the contribution of the first term in the square brackets of (%
\ref{FFb}) becomes zero and one gets%
\begin{equation}
\left\langle F_{\sigma \mu }F_{\rho ^{\prime }\nu ^{\prime }}\right\rangle _{%
\mathrm{b}}=\frac{8\delta _{\mathrm{b}}\alpha ^{3-D}}{\left( 2\pi \right) ^{%
\frac{D}{2}-1}}\eta _{\lbrack \mu \lbrack \rho ^{\prime }}\partial _{\sigma
]}\partial _{\nu ^{\prime }]}\mathcal{J}_{\frac{D}{2}-1}^{(1)}(z,z^{\prime
},s_{+}).  \label{FFb2}
\end{equation}

For the nonzero components with one or two indices being equal to $D$ and
the remaining indices different from $1$ and $D$, we obtain%
\begin{align}
\left\langle F_{D\mu }F_{\rho ^{\prime }\nu ^{\prime }}\right\rangle _{%
\mathrm{b}}& =\frac{4\delta _{\mathrm{b}}\alpha ^{3-D}}{\left( 2\pi \right)
^{\frac{D}{2}-1}}\eta _{\mu \lbrack \rho ^{\prime }}\partial _{\nu ^{\prime
}]}\partial _{z}\mathcal{J}_{\frac{D}{2}-1}^{(1)}(z,z^{\prime },s_{+}), 
\notag \\
\left\langle F_{D\mu }F_{D^{\prime }\nu ^{\prime }}\right\rangle _{\mathrm{b}%
}& =\frac{2\delta _{\mathrm{b}}\alpha ^{3-D}}{\left( 2\pi \right) ^{\frac{D}{%
2}-1}}\left( \partial _{\mu }\partial _{\nu ^{\prime }}u_{+}-\eta _{\mu \nu
^{\prime }}\partial _{z}\partial _{z^{\prime }}\right) \mathcal{J}_{\frac{D}{%
2}-1}^{(1)}(z,z^{\prime },s_{+}),  \label{FF1}
\end{align}%
where $\mu ,\rho ,\nu \neq 1,D$. The components with one of their indices in
the direction of the axis $x^{1}$ become%
\begin{align}
\left\langle F_{1\mu }F_{\rho ^{\prime }\nu ^{\prime }}\right\rangle _{%
\mathrm{b}}& =\frac{4\delta _{\mathrm{b}}\alpha ^{3-D}x_{+}^{1}}{\left( 2\pi
\right) ^{\frac{D}{2}-1}zz^{\prime }}\eta _{\mu \lbrack \nu ^{\prime
}}\partial _{\rho ^{\prime }]}u_{+}\mathcal{J}_{\frac{D}{2}%
}^{(1)}(z,z^{\prime },s_{+}),  \notag \\
\left\langle F_{1\mu }F_{D^{\prime }\nu ^{\prime }}\right\rangle _{\mathrm{b}%
}& =-\frac{2\delta _{\mathrm{b}}\alpha ^{3-D}}{\left( 2\pi \right) ^{\frac{D%
}{2}-1}}\eta _{\mu \nu ^{\prime }}\partial _{1}\partial _{z^{\prime }}%
\mathcal{J}_{\frac{D}{2}-1}^{(1)}(z,z^{\prime },s_{+}),  \notag \\
\left\langle F_{1D}F_{D^{\prime }\nu ^{\prime }}\right\rangle _{\mathrm{b}}&
=-\frac{2\delta _{\mathrm{b}}\alpha ^{3-D}}{\left( 2\pi \right) ^{\frac{D}{2}%
-1}}\partial _{1}\partial _{\nu }^{\prime }u_{+}\mathcal{J}_{\frac{D}{2}%
-1}^{(1)}(z,z^{\prime },s_{+}),  \label{FF2}
\end{align}%
with $\mu ,\rho ,\nu \neq 1,D$. The expressions for the components with two
indices along $x^{1}$ and with $\mu \neq 1$, $\nu \neq 1,D$ read%
\begin{align}
\left\langle F_{1\mu }F_{1^{\prime }\nu ^{\prime }}\right\rangle _{\mathrm{b}%
}& =-\frac{2\delta _{\mathrm{b}}\alpha ^{3-D}}{\left( 2\pi \right) ^{\frac{D%
}{2}-1}}\left[ \partial _{\mu }\partial _{\nu ^{\prime }}+\eta _{\mu \nu
^{\prime }}\partial _{1}\partial _{1^{\prime }}\right] \mathcal{J}_{\frac{D}{%
2}-1}^{(1)}(z,z^{\prime },s_{+}),  \notag \\
\left\langle F_{1D}F_{1^{\prime }D^{\prime }}\right\rangle _{\mathrm{b}}& =-%
\frac{2\delta _{\mathrm{b}}\alpha ^{3-D}}{\left( 2\pi \right) ^{\frac{D}{2}%
-1}}\left\{ zz^{\prime }\partial _{1}\partial _{1}^{\prime }\left[ \frac{%
\partial _{s_{+}}}{s_{+}}\mathcal{J}_{\frac{D}{2}-2}^{(1)}(z,z^{\prime
},s_{+})\right] +\partial _{z}\partial _{z^{\prime }}\mathcal{J}_{\frac{D}{2}%
-1}^{(1)}(z,z^{\prime },s_{+})\right\} .  \label{FF3}
\end{align}%
When deriving the expressions for the field tensor correlators we used the
relation%
\begin{equation}
\partial _{z}\partial _{z^{\prime }}\mathcal{J}_{\nu }^{(-1)}(z,z^{\prime
},s_{+})=-zz^{\prime }\frac{\partial _{s_{+}}}{s_{+}}\mathcal{J}_{\nu
-1}^{(1)}(z,z^{\prime },s_{+}),  \label{Jmin}
\end{equation}%
and the formula (\ref{Jcalder}) for the derivative in the right-hand side.
The relation (\ref{Jmin}) is obtained by using the expressions for
derivatives from (\ref{jfder}). It is important to note here that only the
function $\mathcal{J}_{\nu }^{(1)}(z,z^{\prime },s_{+})$, for which we have
the simple expression (\ref{Jcal1}), is included in the formulas (\ref{FF1}%
)-(\ref{FF3}). For the derivatives we have%
\begin{equation}
\partial _{\mu }\mathcal{J}_{\nu }^{(1)}(z,z^{\prime },s_{+})=\eta _{\mu
\alpha }\frac{\Delta x^{\alpha }}{zz^{\prime }}u_{+}\mathcal{J}_{\nu
+1}^{(1)}(z,z^{\prime },s_{+}),  \label{Jcald2}
\end{equation}%
with $\mu \neq 1,D$. Note that the terms in (\ref{AA4}) with the functions $%
\mathcal{J}_{\frac{D}{2}-1}^{(-1)}(z,z^{\prime },s_{+})$ and $\mathcal{J}_{%
\frac{D}{2}-1}^{(1)}(z,z^{\prime },s_{+})$ both contribute to the Wightman
function of the field tensor. The contribution coming from the function $%
\mathcal{J}_{\frac{D}{2}-1}^{(-1)}(z,z^{\prime },s_{+})$ is expressed in
terms of the function (\ref{Jcal1}) using the formula (\ref{Jmin}). In
particular, it follows from here that the term in (\ref{AA4}) involving $%
\mathcal{J}_{\frac{D}{2}-1}^{(-1)}(z,z^{\prime },s_{+})$ cannot be
eliminated by a gauge transformation.

The two-point functions for a boundary in the Minkowski spacetime are
obtained from the formulas given above in the limit $\alpha \rightarrow
\infty $ for fixed values of the coordinate $y$. By taking into account the
relation $z=\alpha e^{y/\alpha }$, we see that in this limit the coordinate $%
z$ takes large values. For the function appearing in the expressions of the
vacuum correlators we have%
\begin{equation}
\lim_{\alpha \rightarrow \infty }u_{+}^{n}\frac{\mathcal{J}_{\nu
}^{(1)}(z,z^{\prime },s_{+})}{\alpha ^{2\nu -1}}=\frac{2^{\nu -1}\Gamma
\left( \nu +\frac{1}{2}\right) }{\sqrt{\pi }\left[ s_{+}^{2}+\left( \Delta
y\right) ^{2}\right] ^{\nu +\frac{1}{2}}},  \label{limI}
\end{equation}%
with $\Delta y=y-y^{\prime }$ and $n=0,1$. The components with $\mu ,\nu
,\rho \neq 1$ are presented as%
\begin{equation}
\left\langle F_{\sigma \mu }F_{\rho ^{\prime }\nu ^{\prime }}\right\rangle _{%
\mathrm{(M)b}}=4\delta _{\mathrm{b}}\frac{\Gamma \left( \frac{D-1}{2}\right) 
}{\pi ^{\frac{D-1}{2}}}\eta _{\lbrack \mu \lbrack \rho ^{\prime }}\partial
_{\sigma ]}\partial _{\nu ^{\prime }]}\left[ s_{+}^{2}+\left( \Delta
y\right) ^{2}\right] ^{\frac{1-D}{2}}.  \label{FFM1}
\end{equation}%
For the remaining components we get%
\begin{equation}
\left\langle F_{1\mu }F_{1^{\prime }\nu ^{\prime }}\right\rangle _{\mathrm{%
(M)b}}=-\delta _{\mathrm{b}}\frac{\Gamma \left( \frac{D-1}{2}\right) }{\pi ^{%
\frac{D-1}{2}}}\left( \partial _{\mu }\partial _{\nu ^{\prime }}+\eta _{\mu
\nu ^{\prime }}\partial _{1}\partial _{1^{\prime }}\right) \left[
s_{+}^{2}+\left( \Delta y\right) ^{2}\right] ^{\frac{1-D}{2}}.  \label{FFM2}
\end{equation}

The two-point functions given above describe the influence of the boundary
on the correlations of the vacuum fluctuations at different spacetime
points. They can be used for evaluating physical quantities characterizing
the electromagnetic vacuum. For example, the components $\left\langle
F_{0j}F_{0^{\prime }l^{\prime }}\right\rangle _{\mathrm{b}}=\left\langle
E_{j}E_{l^{\prime }}\right\rangle _{\mathrm{b}}$ determine the correlations
of the electric field $E_{j}(x)$ and the Casimir-Polder forces acting on
polarizable particles. In the static limit, when the dispersion of the
polarizability tensor $\alpha _{\mathrm{P}}^{jl}$ can be ignored, the
effective potential for the Casimir-Polder interaction between a particle
and boundary is given by $U_{\mathrm{CP}}=-\alpha _{\mathrm{P}%
}^{jl}\left\langle E_{j}E_{l}\right\rangle _{\mathrm{b}}$. In a more general
case of frequency dependent polarizability tensor $\alpha _{\mathrm{P}%
}^{jl}=\alpha _{\mathrm{P}}^{jl}(\omega )$, the Casimir-Polder potential is
expressed in terms of the integral over imaginary frequencies $\omega =i\xi $
which contains the spectral components of the function $\left\langle
E_{j}E_{l}\right\rangle _{\mathrm{b}}$. These spectral components are found
by the Fourier transformation with respect to $\Delta t=t-t^{\prime }$. The
same two-point functions $\left\langle E_{j}E_{l^{\prime }}\right\rangle _{%
\mathrm{b}}$ determine the influence of the boundary on the response of the
Unruh-DeWitt detector with a dipole coupling to the vacuum fluctuations of
the electromagnetic field. For the detector worldline $x=x(\tau )$
parametrized by proper time $\tau $, the corresponding interaction
Hamiltonian has the form $H_{\mathrm{int}}(\tau )=-d^{l}(\tau )E_{l}(x(\tau
))$, where $d^{l}(\tau )$ is the dipole moment operator. The transition rate
per unit proper time is determined by the integral $\int_{-\infty }^{\infty
}d\Delta \tau \,e^{-i\omega \Delta \tau }\,\left\langle E_{j}(x(\tau
))E_{l}(x(\tau ^{\prime }))\right\rangle $ with $\Delta \tau =\tau -\tau
^{\prime }$.

In the discussion below, as important local characteristics of the
electromagnetic vacuum, we consider the VEVs of the field squares and
energy-momentum tensor.

\section{VEVs of the fields squares and photon condensate}

\label{sec:E2B2}

Given the Wightman function for the field tensor, we can evaluate the VEVs
of local physical quantities, that characterize the electromagnetic vacuum,
by taking the coincidence limit of the arguments. These expectation values
are separated into two contributions. The first corresponds to the VEV of
the problem without a brane, while the second corresponds to the part due to
the presence of a brane. The coincidence limit of two-point functions is
known to be divergent, so renormalization is necessary to obtain finite
values of physical quantities. The structure of divergences in local
physical quantities at a given spacetime point is determined by the
geometric characteristics of the background spacetime at that point, such as
the Riemann tensor and various combinations constructed from it (see, for
example, \cite{Birr82,Park09}). For points not on the brane, these
characteristics are the same in problems with and without branes.
Consequently, the divergences in these problems are also the same at those
points. Therefore, the coincidence limit of the boundary-induced
contributions in the two-point functions is finite for points away from the
brane, and the renormalization of the total VEVs reduces to the
renormalization of the brane-free VEVs.

First we consider the VEV of the electric field squared. It is decomposed as 
$\left\langle E^{2}\right\rangle =\left\langle E^{2}\right\rangle
_{0}+\left\langle E^{2}\right\rangle _{\mathrm{b}}$, where $\left\langle
E^{2}\right\rangle _{0}$ is the VEV in the brane-free geometry. For the
brane-induced part we have 
\begin{equation}
\left\langle E^{2}\right\rangle _{\mathrm{b}}=-\left\langle F_{0\mu }F^{0\mu
}\right\rangle _{\mathrm{b}}.  \label{E2}
\end{equation}%
For $x^{1}\neq 0$, the renormalization is required only for the contribution 
$\left\langle E^{2}\right\rangle _{0}$. From the maximal symmetry of the AdS
spacetime we expect that the corresponding renormalized VEV will not depend
on the spacetime point. By taking into account the expressions for the
correlators of the field tensor given above, from (\ref{E2}) we find%
\begin{equation}
\left\langle E^{2}\right\rangle _{\mathrm{b}}=-\delta _{\mathrm{b}}\Gamma
\left( \frac{D+1}{2}\right) \frac{4(D-2)w^{4}+4(2D-3)w^{2}+3\left(
D-1\right) }{2^{D}\pi ^{\frac{D-1}{2}}\left( \alpha w\right) ^{D+1}\left(
1+w^{2}\right) ^{\frac{D+1}{2}}}.  \label{E2b}
\end{equation}%
Here, we have introduced the notation $w=x^{1}/z$. It measures the proper
distance from the boundary in units of the curvature radius $\alpha $
(scaled proper distance). The brane-induced VEV of the electric field
squared is negative/positive in spatial dimensions $D\geq 2$ for the PMC/PEC
boundary conditions. The Casimir-Polder potential in the case of isotropic
polarizability $\alpha _{\mathrm{P}}$ is given by $U_{\mathrm{CP}}=-\alpha _{%
\mathrm{P}}\left\langle E^{2}\right\rangle _{\mathrm{b}}$. The corresponding
forces are repulsive/attractive for the PMC/PEC boundary conditions. For $%
D=3 $ one finds 
\begin{equation}
\left\langle E^{2}\right\rangle _{\mathrm{b}}=\frac{\delta _{\mathrm{b}}}{%
4\pi \alpha ^{4}}\left[ \frac{1}{\left( 1+w^{2}\right) ^{2}}-\frac{3}{w^{4}}%
\right] .  \label{E2bD3}
\end{equation}

Next we consider the brane-induced contribution in the photon condensate
defined as the VEV $\left\langle F_{\sigma \mu }F^{\sigma \mu }\right\rangle
_{\mathrm{b}}$. It is an analog of the gluon condensate in quantum
chromodynamics. Taking into account the expressions (\ref{FFb2})-(\ref{FF3})
for the components of the Wightman function, one obtains%
\begin{equation}
\left\langle F_{\sigma \mu }F^{\sigma \mu }\right\rangle _{\mathrm{b}%
}=\delta _{\mathrm{b}}\frac{D\Gamma \left( \frac{D+1}{2}\right) }{2^{D-2}\pi
^{\frac{D-1}{2}}}\frac{2(D-2)w^{2}+D-1}{\left( \alpha w\right) ^{D+1}\left(
1+w^{2}\right) ^{\frac{D-1}{2}}}.  \label{Lb2}
\end{equation}%
This VEV is negative/positive for PMC/PEC conditions. For the magnetic field
squared we have%
\begin{equation}
\left\langle B^{2}\right\rangle _{\mathrm{b}}=\left\langle
E^{2}\right\rangle _{\mathrm{b}}+\frac{1}{2}\left\langle F_{\sigma \mu
}F^{\sigma \mu }\right\rangle _{\mathrm{b}}.  \label{B2b}
\end{equation}%
By taking into account (\ref{E2b}) and (\ref{Lb2}), one gets%
\begin{equation}
\left\langle B^{2}\right\rangle _{\mathrm{b}}=\delta _{\mathrm{b}%
}(D-1)\Gamma \left( \frac{D+1}{2}\right) \frac{4(D-2)w^{4}+6(D-2)w^{2}+2D-3}{%
2^{D}\pi ^{\frac{D-1}{2}}\left( \alpha w\right) ^{D+1}\left( 1+w^{2}\right)
^{\frac{D+1}{2}}}.  \label{B2b2}
\end{equation}%
The boundary-induced VEV of the magnetic field squared is positive/negative
for PMC/PEC boundary conditions. For $D=3$ we obtain the following simple
expression:%
\begin{equation}
\left\langle B^{2}\right\rangle _{\mathrm{b}}=\frac{\delta _{\mathrm{b}}}{%
4\pi \alpha ^{4}}\left[ \frac{1}{\left( 1+w^{2}\right) ^{2}}+\frac{3}{w^{4}}%
\right] .  \label{B2bD3}
\end{equation}%
The VEVs $\left\langle E^{2}\right\rangle _{\mathrm{b}}$ and $\left\langle
B^{2}\right\rangle _{\mathrm{b}}$ are symmetric with respect to the boundary 
$x^{1}=0$. The expressions in the region $x^{1}<0$ are obtained from the
formulas above by the replacement $x^{1}\rightarrow |x^{1}|$.

Let us consider the behavior of the VEVs in the asymptotic regions of the
variable $w$ being the proper distance from the boundary in units of $\alpha 
$. Near the boundary we have $w\ll 1$ and the leading terms in the
expansions of the brane-induced VEVs are given by%
\begin{equation}
\left\langle E^{2}\right\rangle _{\mathrm{b}}\approx -\frac{\left\langle
B^{2}\right\rangle _{\mathrm{b}}}{2D/3-1}\approx -\frac{3\delta _{\mathrm{b}%
}\left( D-1\right) \Gamma \left( \frac{D+1}{2}\right) }{2^{D}\pi ^{\frac{D-1%
}{2}}\left( \alpha w\right) ^{D+1}}.  \label{EBnear}
\end{equation}%
These asymptotes also describe the behavior of the VEVs near the AdS horizon
(large values of $z$ for fixed $x^{1}$). At large distances from the
boundary, corresponding to $w\gg 1$, the decay of the VEVs is described by%
\begin{equation}
\left\langle E^{2}\right\rangle _{\mathrm{b}}\approx -\frac{\left\langle
B^{2}\right\rangle _{\mathrm{b}}}{D-1}\approx -\frac{4\delta _{\mathrm{b}%
}(D-2)\Gamma \left( \frac{D+1}{2}\right) }{2^{D}\pi ^{\frac{D-1}{2}}\alpha
^{D+1}w^{2D-2}},  \label{EBfar}
\end{equation}%
for $D>2$. In the case $D=2$, the large distance behavior is given by 
\begin{equation}
\left\langle E^{2}\right\rangle _{\mathrm{b}}\approx -\frac{\delta _{\mathrm{%
b}}}{2\alpha ^{3}w^{4}},\;\left\langle B^{2}\right\rangle _{\mathrm{b}%
}\approx \frac{\delta _{\mathrm{b}}}{8\alpha ^{3}w^{6}}.  \label{EBfarD2}
\end{equation}%
For a given $x^{1}$, the asymptotics (\ref{EBfar}) and (\ref{EBfarD2})
describe the behavior of the VEVs near the AdS boundary.

The VEVs for a boundary in the Minkowski spacetime are obtained from the
formulas given above in the limit $\alpha \rightarrow \infty $ and $\alpha
w\rightarrow x^{1}$. The corresponding expressions read 
\begin{equation}
\left\langle E^{2}\right\rangle _{\mathrm{(M)}}=-\frac{\left\langle
B^{2}\right\rangle _{\mathrm{(M)}}}{2D/3-1}=-\frac{3\delta _{\mathrm{b}%
}\left( D-1\right) \Gamma \left( \frac{D+1}{2}\right) }{2^{D}\pi ^{\frac{D-1%
}{2}}(x^{1})^{D+1}}.  \label{EBMink}
\end{equation}%
Note that the leading terms in the asymptotic expansions of the VEVs near
the boundary, given by (\ref{EBnear}), are given by the same expressions
with the distance from the boundary replaced by the proper distance $\alpha
x^{1}/z$ in the problem on the AdS bulk. This shows that in the region under
consideration the effects of the curvature are weak. This is because the
contribution of short-wavelength vacuum modes is dominant at points near the
boundary. These modes are weakly influenced by the gravitational field. At
distances larger than the curvature radius of the background geometry, the
influence of the gravitational field is essential. The suppression of the
boundary-induced VEVs at large distances, as functions of the proper
distance from the boundary, is stronger in the AdS bulk in spatial
dimensions $D\neq 3$. For $D=3$, the VEVs decay as the fourth power of the
proper distance in both cases of the Minkowski and AdS bulks. This is
related to the conformal invariance of the electromagnetic field in
(3+1)-dimensional spacetime.

\begin{figure}[tbph]
\begin{center}
\epsfig{figure=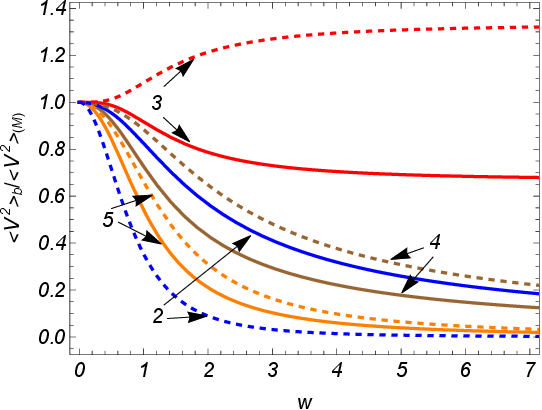,width=8cm,height=6.5cm}
\end{center}
\caption{The ratios of the boundary-induced VEVs in the electric ($V=E$,
solid curves) and magnetic ($V=B$, dashed curves) fields squares on the AdS
and Minkowski bulks versus the proper distance from the boundary in units of 
$\protect\alpha $. The numbers near the curves correspond to the values of
the spatial dimension.}
\label{fig1}
\end{figure}

In Fig. \ref{fig1} we have plotted the ratio of the boundary-induced VEVs of
the electric and magnetic fields squares in the problems of the AdS and
Minkowski bulks, as functions of the proper distance from the brane in units
of $\alpha $. The latter is given by $w=x^{1}/\alpha $ in the Minkowski
spacetime and by $w=x^{1}/z$ for the AdS bulk. The solid and dashed curves
correspond to the electric ($V=E$) and magnetic ($V=B$) fields and the
numbers present the values of the spatial dimension $D$. We recall that the
VEVs $\left\langle E^{2}\right\rangle _{\mathrm{b}}$ and $\left\langle
B^{2}\right\rangle _{\mathrm{b}}$ have opposite signs. One has $\left\langle
E^{2}\right\rangle _{\mathrm{b}}>0$ for PEC condition and $\left\langle
E^{2}\right\rangle _{\mathrm{b}}<0$ for PMC condition. Note that for $D\neq
3 $ the suppression of the boundary-induced VEVs in the AdS bulk is stronger
compared to the corresponding VEVs in the Minkowski spacetime. For $D=3$ we
have the limiting values $\left\langle E^{2}\right\rangle _{\mathrm{b}%
}/\left\langle E^{2}\right\rangle _{\mathrm{(M)}}\rightarrow 2/3$ and $%
\left\langle B^{2}\right\rangle _{\mathrm{b}}/\left\langle
B^{2}\right\rangle _{\mathrm{(M)}}\rightarrow 4/3$ for $w\rightarrow \infty $%
. 

\section{Energy-momentum tensor}

\label{sec:EMT}

Another important local characteristic of the vacuum state is the VEV of the
energy-momentum tensor:%
\begin{equation}
\langle T_{\nu }^{\mu }\rangle =-\frac{\langle F_{\nu \rho }F^{\mu \rho
}\rangle }{4\pi }+\delta _{\nu }^{\mu }\frac{\left\langle F_{\sigma \rho
}F^{\sigma \rho }\right\rangle }{16\pi }.  \label{Tmunu}
\end{equation}%
Similar to the VEVs in the previous Section, it is a sum of the brane-free
and brane-induced parts, denoted here by $\langle T_{\nu }^{\mu }\rangle
_{0} $ and $\langle T_{\nu }^{\mu }\rangle _{\mathrm{b}}$, respectively.
From the maximal symmetry of the AdS spacetime one has $\langle T_{\nu
}^{\mu }\rangle _{0}=\mathrm{const}\cdot \delta _{\nu }^{\mu }$ for the
boundary-free contribution. For points with $x^{1}\neq 0$, the divergences
are removed by the renormalization of $\langle T_{\nu }^{\mu }\rangle _{0}$.
By using the expressions (\ref{FFb2})-(\ref{FF3}) for the two-point
functions, the brane-induced contributions in the VEVs of the components
with $\ \mu \neq 1,D$ are presented in the form (no summation over $\mu $)%
\begin{equation}
\langle T_{\mu }^{\mu }\rangle _{\mathrm{b}}=\delta _{\mathrm{b}}\Gamma
\left( \frac{D+1}{2}\right) \frac{2(D-2)^{2}w^{4}+(D-3)\left[ (3D-4)w^{2}+D-1%
\right] }{2^{D+2}\pi ^{\frac{D+1}{2}}\left( \alpha w\right) ^{D+1}\left(
1+w^{2}\right) ^{\frac{D+1}{2}}}.  \label{Tmu}
\end{equation}%
For the remaining diagonal stresses one gets%
\begin{align}
\langle T_{1}^{1}\rangle _{\mathrm{b}}& =\delta _{\mathrm{b}}\Gamma \left( 
\frac{D+1}{2}\right) \frac{2(D-2)^{2}w^{4}+\left( D^{2}-9D+12\right)
w^{2}+3-D}{2^{D+2}\pi ^{\frac{D+1}{2}}\alpha ^{D+1}w^{D-1}\left(
1+w^{2}\right) ^{\frac{D+3}{2}}},  \notag \\
\langle T_{D}^{D}\rangle _{\mathrm{b}}& =\delta _{\mathrm{b}}\Gamma \left( 
\frac{D+1}{2}\right) \frac{2D(2-D)w^{6}+(D^{2}-5D+8)w^{4}+(D-3)\left[
(3D-4)w^{2}+D-1\right] }{2^{D+2}\pi ^{\frac{D+1}{2}}\left( \alpha w\right)
^{D+1}\left( 1+w^{2}\right) ^{\frac{D+3}{2}}}.  \label{TDD}
\end{align}%
We can see that the all off-diagonal components of the energy-momentum
tensor vanish except the stress $\langle T_{1}^{D}\rangle _{\mathrm{b}%
}=\langle T_{D}^{1}\rangle _{\mathrm{b}}$. The corresponding expression
becomes%
\begin{equation}
\langle T_{1}^{D}\rangle _{\mathrm{b}}=\delta _{\mathrm{b}}(D-1)\Gamma
\left( \frac{D+1}{2}\right) \frac{4(D-2)w^{4}+(D-3)\left( 4w^{2}+1\right) }{%
2^{D+2}\pi ^{\frac{D+1}{2}}\alpha ^{D+1}w^{D}\left( 1+w^{2}\right) ^{\frac{%
D+3}{2}}}.  \label{TD1}
\end{equation}%
In spatial dimensions $D\geq 3$, the energy density, given by (\ref{Tmu})
with $\mu =0$, is positive/negative for PMC/PEC boundary conditions. For $%
D=2 $, the energy density becomes negative/positive for PMC/PEC conditions.
The expressions (\ref{Tmu})-(\ref{TD1}) are given for the region $x^{1}>0$.
The diagonal components in the region $x^{1}<0$ are expressed by the same
formulas with the replacement $x^{1}\rightarrow |x^{1}|$. For the
off-diagonal component (\ref{TD1}) in addition the sign should be changed.

In problems involving quantum fields on the Poincar\'{e} patch of an AdS
bulk, in the absence of branes, the geometry remains invariant under Lorentz
boosts that are parallel to the AdS boundary. This invariance is also
preserved in problems with branes parallel to the AdS boundary, provided
that Lorentz-invariant boundary conditions are imposed on the quantum
fields. This Lorentz invariance implies that the part of the vacuum
energy-momentum tensor with indices in the subspace parallel to the AdS
boundary is diagonal, and $\langle T_{\nu }^{\mu }\rangle =\delta _{\nu
}^{\mu }\langle T_{0}^{0}\rangle $ for $\mu ,\nu =1,2,\ldots ,D-1$. In the
problem under consideration, the presence of a brane orthogonal to the AdS
boundary breaks the homogeneity along the axis $x^{1}$. The appearance of
the non-zero off-diagonal component (\ref{TD1}) is conditioned by this
inhomogeneity, combined with the dependence of the metric tensor on the $z$%
-coordinate.

As an additional check, it can be seen that the brane-induced parts of the
vacuum energy-momentum tensor obey the covariant continuity equation $\nabla
_{\mu }\left\langle T_{\rho }^{\mu }\right\rangle _{\mathrm{b}}=0$. For the
problem under consideration it reduces to the following two equations:%
\begin{equation}
\partial _{1}\left\langle T_{1}^{1}\right\rangle _{\mathrm{b}%
}+z^{D+1}\partial _{z}\left( z^{-D-1}\left\langle T_{1}^{D}\right\rangle _{%
\mathrm{b}}\right) =0,  \label{Conteq1}
\end{equation}%
and%
\begin{equation}
\partial _{1}\left\langle T_{D}^{1}\right\rangle _{\mathrm{b}}+z^{D}\partial
_{z}\left( z^{-D}\left\langle T_{D}^{D}\right\rangle _{\mathrm{b}}\right)
+z^{-1}\left[ (D-1)\left\langle T_{0}^{0}\right\rangle _{\mathrm{b}%
}+\left\langle T_{1}^{1}\right\rangle _{\mathrm{b}}\right] =0.
\label{Conteq2}
\end{equation}%
By taking into account that the VEVs are functions of $w$, these relations
can also be written in the form%
\begin{align}
0& =\partial _{w}\left\langle T_{1}^{1}\right\rangle _{\mathrm{b}%
}-w^{-D}\partial _{w}\left( w^{D+1}\left\langle T_{1}^{D}\right\rangle _{%
\mathrm{b}}\right) ,  \notag \\
0& =\partial _{w}\left\langle T_{D}^{1}\right\rangle _{\mathrm{b}%
}-w^{1-D}\partial _{w}\left( w^{D}\left\langle T_{D}^{D}\right\rangle _{%
\mathrm{b}}\right) +(D-1)\left\langle T_{0}^{0}\right\rangle _{\mathrm{b}%
}+\left\langle T_{1}^{1}\right\rangle _{\mathrm{b}}.  \label{Conteq3}
\end{align}%
In addition, we have the trace relation $\langle T_{\mu }^{\mu }\rangle _{%
\mathrm{b}}=(D-3)\left\langle F_{\sigma \mu }F^{\sigma \mu }\right\rangle _{%
\mathrm{b}}/16\pi $.

\begin{figure}[tbph]
\begin{center}
\epsfig{figure=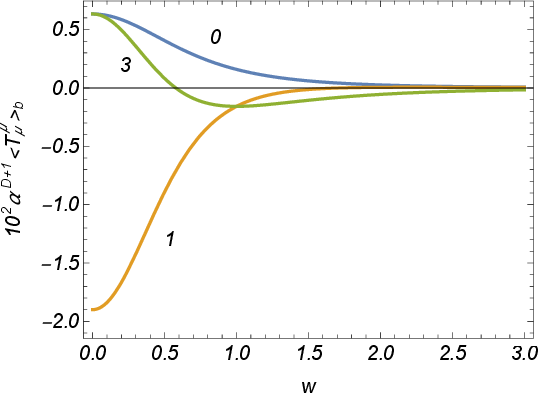,width=8cm,height=6.5cm}
\end{center}
\caption{Brane-induced diagonal components of the vacuum energy-momentum
tensor in spatial dimension $D=3$ versus the proper distance from the
boundary. The graphs are plotted for the PMC condition and the numbers near
the curves correspond to the value of the index $\protect\mu $. }
\label{fig2}
\end{figure}

For $D=3$, one gets 
\begin{align}
\langle T_{0}^{0}\rangle _{\mathrm{b}}& =\langle T_{2}^{2}\rangle _{\mathrm{b%
}}=\delta _{\mathrm{b}}\frac{\left( 1+w^{2}\right) ^{-2}}{16\pi ^{2}\alpha
^{4}}\ ,  \notag \\
\langle T_{1}^{1}\rangle _{\mathrm{b}}& =\frac{\delta _{\mathrm{b}}}{16\pi
^{2}\alpha ^{4}}\frac{w^{2}-3}{(1+w^{2})^{3}},  \notag \\
\left\langle T_{3}^{3}\right\rangle _{\mathrm{b}}& =\frac{\delta _{\mathrm{b}%
}}{16\pi ^{2}\alpha ^{4}}\frac{1-3w^{2}}{\left( 1+w^{2}\right) ^{3}},
\label{TD3}
\end{align}%
for the diagonal components and 
\begin{equation}
\left\langle T_{1}^{D}\right\rangle _{\mathrm{b}}=\frac{\delta _{\mathrm{b}}%
}{4\pi ^{2}\alpha ^{4}}\frac{w}{\left( 1+w^{2}\right) ^{3}},  \label{T1D3}
\end{equation}%
for the off-diagonal component. In this special case, the brane-induced VEV
of the energy-momentum tensor is traceless. The trace anomaly is contained
in the brane-free VEV. The VEVs (\ref{TD3}) are finite on the brane with (no
summation over $\mu =0,2,3$) $\langle T_{\mu }^{\mu }\rangle _{\mathrm{b}%
}=-\langle T_{1}^{1}\rangle _{\mathrm{b}}/3=\delta _{\mathrm{b}}/(16\pi
^{2}\alpha ^{4})$ and the off-diagonal component becomes zero. The latter
means that the force acting on the boundary along the $z$-direction (shear
force) vanishes. In Fig. \ref{fig2}, the diagonal components $\langle T_{\mu
}^{\mu }\rangle _{\mathrm{b}}$ from (\ref{TD3}) are plotted for the PMC
condition as functions of the proper distance from the boundary (for the
off-diagonal component see Fig. \ref{fig4}). The numbers near the curves are
the values of $\mu $. 

As it has been mentioned above, for $D=3$ the problem under consideration is
conformally related to the problem in the Minkowski bulk with two planar
boundaries located at $x^{1}=0$ and $z=0$. On the latter boundary the field
obeys PEC condition. The corresponding VEV $\langle T_{\mu }^{\nu }\rangle _{%
\mathrm{(M)}}$ is obtained from the expressions on the AdS bulk by using the
conformal relation $\langle T_{\mu }^{\nu }\rangle _{\mathrm{(M)}}=(z/\alpha
)^{4}\langle T_{\mu }^{\nu }\rangle _{\mathrm{b}}$. This gives:%
\begin{equation}
\langle T_{\mu }^{\nu }\rangle _{\mathrm{(M)}}=\frac{\delta _{\mathrm{b}%
}g_{\mu }^{\nu }(x^{1},z)}{16\pi ^{2}[\left( x^{1}\right) ^{2}+z^{2}]^{3}},
\label{TMink2}
\end{equation}%
with the notations%
\begin{equation}
g_{0}^{0}(u,z)=g_{2}^{2}(u,z)=u^{2}+z^{2},\;g_{1}^{1}(u,z)=u^{2}-3z^{2},%
\;g_{3}^{3}(u,z)=z^{2}-3u^{2},  \label{g00}
\end{equation}%
and $g_{1}^{D}(u,z)=4uz$. The remaining components of the function $g_{\mu
}^{\nu }(u,z)$ become zero.

For general $D$, near the brane, $w\ll 1$, the asymptotics of the
brane-induced mean energy-momentum tensor components are given by (no
summation over $\mu $)%
\begin{equation}
\langle T_{\mu }^{\mu }\rangle _{\mathrm{b}}\approx \frac{1-D}{w^{2}}\langle
T_{1}^{1}\rangle _{\mathrm{b}}\approx \frac{1}{w}\langle T_{1}^{D}\rangle _{%
\mathrm{b}}\approx \delta _{\mathrm{b}}\frac{(D-3)\left( D-1\right) \Gamma
\left( \frac{D+1}{2}\right) }{2^{D+2}\pi ^{\frac{D+1}{2}}\left( \alpha
w\right) ^{D+1}},  \label{Tmunear}
\end{equation}%
where $\mu \neq 1,D$. In the special case $D=3$, the vacuum energy-momentum
tensor is finite on the brane (no summation over $\mu $):%
\begin{equation}
\langle T_{\mu }^{\mu }\rangle _{\mathrm{b}}\approx -\frac{1}{3}\langle
T_{1}^{1}\rangle _{\mathrm{b}}=\delta _{\mathrm{b}}\frac{1+\mathcal{O}(w^{2})%
}{16\pi ^{2}\alpha ^{4}},\;\left\langle T_{1}^{D}\right\rangle _{\mathrm{b}%
}\approx \frac{\delta _{\mathrm{b}}w}{4\pi ^{2}\alpha ^{4}}.
\label{TmuNearD3}
\end{equation}%
At large distances, $w\gg 1$, we get%
\begin{equation}
\langle T_{\mu }^{\mu }\rangle _{\mathrm{b}}\approx \frac{2-D}{D}\langle
T_{D}^{D}\rangle _{\mathrm{b}}\approx \frac{w}{2}\frac{D-2}{D-1}\langle
T_{1}^{D}\rangle _{\mathrm{b}}\approx \delta _{\mathrm{b}}\frac{%
(D-2)^{2}\Gamma \left( \frac{D+1}{2}\right) }{2^{D+1}\pi ^{\frac{D+1}{2}%
}\alpha ^{D+1}w^{2D-2}},  \label{Tmufar}
\end{equation}%
with $\mu \neq D$. In the special case $D=2$, the leading behavior at large
distances is given by%
\begin{equation}
\langle T_{\mu }^{\mu }\rangle _{\mathrm{b}}\approx -\langle
T_{D}^{D}\rangle _{\mathrm{b}}\approx \frac{w}{2}\langle T_{1}^{D}\rangle _{%
\mathrm{b}}\approx \frac{-\delta _{\mathrm{b}}}{16\pi \alpha ^{3}w^{4}},
\label{TmufarD2}
\end{equation}%
again, for $\mu \neq D$.

\begin{figure}[tbph]
\begin{center}
\epsfig{figure=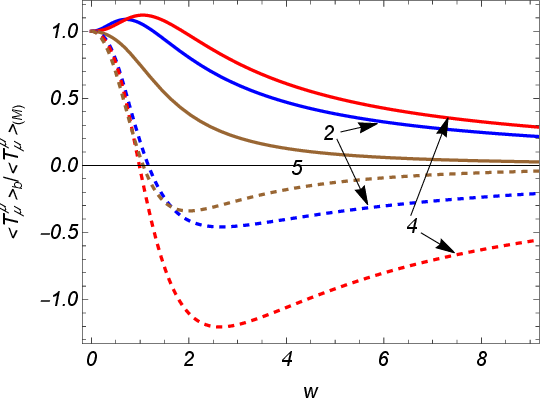,width=8cm,height=6.5cm}
\end{center}
\caption{The ratios of the boundary-induced VEVs in the energy density
(solid curves) and $_{D}^{D}$-stress (dashed curves) on the AdS and
Minkowski bulks versus the scaled proper distance from the boundary. The
numbers near the curves are the values of the spatial dimension.}
\label{fig3}
\end{figure}

\begin{figure}[tbph]
\begin{center}
\begin{tabular}{cc}
\epsfig{figure=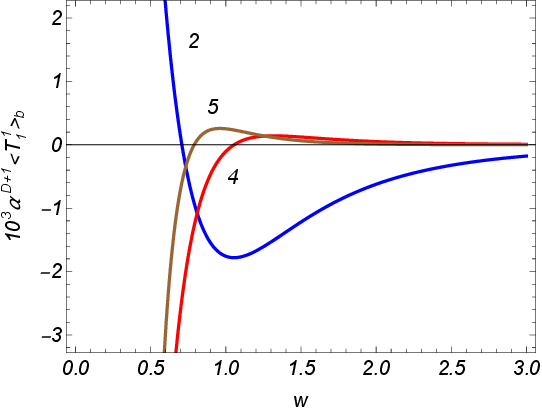,width=8cm,height=6.5cm} & \quad %
\epsfig{figure=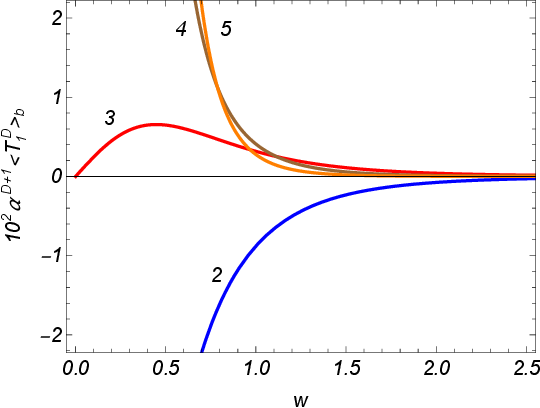,width=8cm,height=6.5cm}%
\end{tabular}%
\end{center}
\caption{Brane-induced contributions to the normal stress (left panel) and
the off-diagonal component of the vacuum energy-momentum tensor (right
panel) for the PMC condition as functions of the distance from the brane.
The numbers near the curves present the values of $D$.}
\label{fig4}
\end{figure}

For a boundary in the Minkowski spacetime the vacuum energy-momentum tensor
is diagonal, $\langle T_{1}^{D}\rangle _{\mathrm{(M)b}}=0$, and (no
summation over $\mu \neq 1$) 
\begin{equation}
\langle T_{\mu }^{\mu }\rangle _{\mathrm{(M)}}=\delta _{\mathrm{b}}\frac{%
(D-3)\left( D-1\right) \Gamma \left( \frac{D+1}{2}\right) }{2^{D+2}\pi ^{%
\frac{D+1}{2}}\left( x^{1}\right) ^{D+1}}.  \label{TmuM}
\end{equation}%
The normal stress in the Minkowskian problem vanishes, $\langle
T_{1}^{1}\rangle _{\mathrm{(M)}}=0$. In spatial dimension $D=3$ we have $%
\langle T_{\nu }^{\mu }\rangle _{\mathrm{(M)}}=0$. For $D\neq 3$, the
leading terms in the near-boundary expansion of the diagonal components for
the AdS bulk are obtained from the VEVs in the Minkowski bulk by the
replacement of the distance from the boundary by the proper distance $\alpha
x^{1}/z$. Figure \ref{fig3} displays the ratio $\langle T_{\mu }^{\mu
}\rangle _{\mathrm{b}}/\langle T_{\mu }^{\mu }\rangle _{\mathrm{(M)}}$ for
the components $\mu =0$ (solid curves) and $\mu =D$ (dashed curves). This
ratio is evaluated for the same values of the proper distance from the
boundary (in units of $\alpha $) in the AdS and Minkowski bulks. The numbers
near the curves present the corresponding values of $D$. 

In Fig. \ref{fig4} we plotted the dependence of the components $\langle
T_{1}^{1}\rangle _{\mathrm{b}}$ (left panel) and $\langle T_{1}^{D}\rangle _{%
\mathrm{b}}$ (right panel) of the vacuum energy-momentum tensor (in units of 
$\alpha ^{-D-1}$) on $w$ for different values of the spatial dimension $D$
(the numbers near the curves) and for the PMC boundary condition. The normal
stress in the spatial dimension $D=3$ is given in Fig. \ref{fig2}. 

It is of interest to compare the VEVs for the electromagnetic field and
scalar field $\varphi (x)$ with mass $m$ and curvature coupling parameter $%
\xi $. The parameters $m$ and $\xi $ enter in the field equation through the
combination $m_{\mathrm{eff}}^{2}=m^{2}+\xi R$, where $R=-D(D+1)/\alpha ^{2}$
is the Ricci scalar for the AdS spacetime. The influence of a brane
orthogonal to the AdS boundary on the local characteristics of scalar vacuum
is discussed in \cite{Beze15} for Robin boundary condition. Here we will
compare the results with those for the Dirichlet and Neumann boundary
conditions. The VEVs of the scalar field squared and energy-momentum tensor
are expressed in terms of the function 
\begin{equation}
g_{\nu }(u)=\frac{\Gamma (\nu +D/2)}{\Gamma (\nu +1)u^{\nu +D/2}}%
\,{}_{2}F_{1}\left( \frac{D+2\nu +2}{4},\frac{D+2\nu }{4};\nu +1;\frac{1}{%
u^{2}}\right) ,  \label{gnu}
\end{equation}%
where $\,{}_{2}F_{1}\left( a,b;c;x\right) $ is the hypergeometric function
and $\nu =\sqrt{D^{2}/4+m_{\mathrm{eff}}^{2}\alpha ^{2}}$.

The brane-induced contribution in the VEV of the field squared is given by 
\cite{Beze15}%
\begin{equation}
\langle \varphi ^{2}\rangle _{\mathrm{b}}=\mp \frac{g_{\nu }(1+2w^{2})}{2^{%
\frac{D}{2}+\nu +1}\pi ^{\frac{D}{2}}\alpha ^{D-1}},  \label{phi2sc}
\end{equation}%
where the upper and lower signs correspond to the Dirichlet and Neumann
conditions respectively. Similarly, the VEVs of the components of the
energy-momentum tensor are expressed in terms of the function (\ref{gnu})
and its first and second derivatives. For example, the VEV of the energy
density in the case of a minimally coupled field ($\xi =0$) reads%
\begin{equation}
\langle T_{0}^{0}\rangle _{\mathrm{b}}^{\mathrm{(sc)}}=\pm \frac{\pi ^{-%
\frac{D}{2}}\alpha ^{-D-1}}{2^{\frac{D}{2}+\nu +1}}\left\{ \left(
1-u^{2}\right) g_{\nu }^{\prime \prime }(u)+\left[ \left( \frac{D}{2}%
+1\right) \left( 1-u\right) -6\right] g_{\nu }^{\prime }(u)\right\} ,
\label{T00sc}
\end{equation}%
with $u=1+2w^{2}$. Note that unlike the case of the electromagnetic field,
the expectation values for a scalar field with zero mass (including the more
important special cases of minimal and conformally coupled fields) are not
expressed in terms of elementary functions.

At large distances from the brane, $w\gg 1$, and for the general case of the
parameter $\xi $, the scalar VEVs decay as $\langle \varphi ^{2}\rangle _{%
\mathrm{b}}\varpropto 1/w^{D+2\nu }$ and (no summation over $\mu $) $\langle
T_{\mu }^{\mu }\rangle _{\mathrm{b}}^{\mathrm{(sc)}}\varpropto 1/w^{D+2\nu }$%
. This decay follows a power law for both massless and massive fields. In
the special case of $\nu =D/2-1$, the power law is the same for the scalar
field and the massless vector field. This corresponds to the condition%
\begin{equation}
m_{\mathrm{eff}}^{2}=(1-D)/\alpha ^{2}  \label{meff}
\end{equation}%
for the effective mass. Of interest is the fact that, for $\nu =D/2-1$, the
function (\ref{gnu}) is expressed in terms of elementary functions:%
\begin{equation}
g_{\frac{D}{2}-1}(u)=\frac{2^{D-2}\Gamma \left( \frac{D-1}{2}\right) }{\sqrt{%
\pi }\left( u^{2}-1\right) ^{\frac{D-1}{2}}}.  \label{gD2}
\end{equation}%
Note that we have the simple relation 
\begin{equation}
\mathcal{J}_{\frac{D}{2}-1}^{(1)}(z,z^{\prime },s_{+})|_{x^{\prime }=x}=2^{-%
\frac{D}{2}}z^{-2}g_{\frac{D}{2}-1}(1+2w^{2}),  \label{Icalg}
\end{equation}%
between the functions entering in the expressions of the VEVs for scalar and
vector fields. This shows that the behavior of the vector field VEVs is most
closely mimicked by the VEV of a scalar field with an effective mass squared
given by (\ref{meff}). The relationship between the gauge boson and scalar
field propagators in AdS spacetime without additional boundaries has been
discussed in \cite{Liu99,Hoke99}. It has been shown that, for a given
current density, the transverse part of the vector potential is expressed in
terms of the propagator for a minimally coupled free scalar field with mass $%
m^{2}=(1-D)/\alpha ^{2}$. This relation coincides with (\ref{meff}) in the
special case $\xi =0$.

Note that, in the setup under consideration, both the bulk and boundary
geometries are fixed. An important area of research would be studying the
backreaction of quantum vacuum effects on the background geometry and brane
dynamics. Unlike classical sources of gravity, the vacuum expectation values
of the energy-momentum tensor can violate the energy conditions in the
singularity theorems, resulting in interesting gravitational dynamics.
Another research direction is the effect of the brane's finite thickness on
the local characteristics of the vacuum. For example, one can consider a
finite-thickness brane whose internal geometry differs from the background
geometry. In this case, the boundary conditions are derived from the
matching conditions for the gravitational and electromagnetic fields on the
surface of the brane. A study of this type for a scalar field in the
geometry of a brane parallel to the AdS boundary is presented in \cite%
{Saha07}.

\section{Conclusion}

\label{sec:Conc}

We discussed the interplay of a brane and background geometry on the
expectation values of local observables for the electromagnetic vacuum in
AdS spacetime. For a brane perpendicular to the AdS boundary, we considered
two types of boundary conditions, which are generalizations of the perfect
magnetic and perfect electric conditions in Maxwell electrodynamics to a
general number of spatial dimensions. The corresponding mode functions for
the vector potential, which are used in the canonical quantization of the
field, are given by (\ref{Amu}). For $D>3$, these modes obey both the PMC
and PEC conditions on the AdS boundary. In three spatial dimensions, the
modes on the AdS boundary are constrained only by the PEC condition. For $%
D=2 $, the mode functions satisfy only the PMC condition. In all these
cases, we have an ideal reflection from the AdS boundary. The properties of
the vacuum state are encoded in two-point functions. As the first step, we
evaluated the Wightman functions of the vector potential and field tensor by
using the summation over the complete set of electromagnetic field modes (%
\ref{Amu}). The brane-induced contributions in these two-point functions are
extracted. The contribution to the Wightman function for the vector
potential is expressed in terms of the integrals (\ref{Jcal}) with $n=\pm 1$%
. For the integral with $n=1$ one has a simple representation (\ref{Jcal1}).
We have shown that all the components of the Wightman function of the field
tensor are expressed in terms of the function $\mathcal{J}_{\nu
}^{(1)}(z,z^{\prime },s_{+})$, as given in (\ref{Jcal1}).

Having all the components of the Wightman function for the field tensor, the
VEVs of the electric and magnetic fields squares and of the energy-momentum
tensor are obtained in the coincidence limit of the spacetime points. For
points away from the brane this limit is finite for the brane-induced
contributions and the renormalization is required only for the brane-free
parts. The brane-induced contributions in the VEVs of the electric and
magnetic fields squares are given by the expressions (\ref{E2b}), (\ref{B2b2}%
), and for the photon condensate the formula (\ref{Lb2}) is obtained. For
PMC/PEC boundary condition, this contribution is negative/positive for the
electric field and positive/negative for magnetic field. The VEVs of the
bilinear products of the electric field components determine the
Casimir-Polder potential for the interaction of a polarizable particle with
the boundary. In the simplest case of isotropic polarizability the
corresponding forces are repulsive/attractive for PMC/PEC condition.

The boundary-induced contributions in the VEVs of the nonzero components of
the energy-momentum tensor are expressed by the formulas (\ref{Tmu})-(\ref%
{TD1}). In addition to the diagonal components one has also an off-diagonal
stress $\langle T_{1}^{D}\rangle _{\mathrm{b}}$. Unlike the problem with a
planar boundary in the Minkowski bulk, the vacuum energy-momentum tensor
does not vanish in (3+1)-dimensional AdS spacetime. In this dimension the
brane-induced VEV is traceless. The trace anomaly is contained in the
boundary-free contribution. Another feature of the problem in 3-dimensional
space is that the components of the vacuum energy-momentum tensor are finite
on the boundary. The corresponding shear force acting along the $z$%
-direction vanishes. In dimensions with $D\geq 3$, the brane-induced vacuum
energy density is positive for PMC condition and negative for PEC condition.
In the region near the brane, for the VEVs of the field squares and diagonal
components of the energy-momentum tensor, the leading terms of the
expansions over the distance from the brane coincide with the corresponding
VEVs in the Minkowski bulk. The effects of the spacetime curvature are
essential for proper distances from the boundary larger than the curvature
radius. At large distances and for AdS bulk with $D\neq 3$, the fall-off of
the brane-induced contributions in the VEVs, as functions of the distance
from the boundary, is stronger compared to the corresponding behavior in the
Minkowski spacetime. Due to the maximal symmetry of the AdS spacetime, the
VEVs depend on the coordinates $x^{1}$ and $z$ through the scaled proper
distance $x^{1}/z$. As a consequence, the asymptotes at large and small
distances from the brane also describe the behavior of the expectation
values near the AdS boundary and horizon. In particular, the VEVs of the
electric and magnetic field squares and of the diagonal components of the
energy-momentum tensor tend to zero on the AdS boundary like $z^{2D-2}$ for $%
D\geq 3$.

The analogue of the PEC and PMC boundary conditions for a scalar field is
the Dirichlet and Neumann conditions. We compared the VEVs obtained in this
paper with the expectation values for a massive scalar field with general
curvature coupling parameter obeying these conditions on a brane
perpendicular to the AdS boundary. Unlike massless vector fields, VEVs for
scalar fields are generally not expressed in terms of elementary functions.
The exception is a scalar field with an effective mass squared given as (\ref%
{meff}). In this case, the large-distance asymptotes of the scalar VEVs
follow the same power-law behavior as the electromagnetic field.

\section*{Acknowledgments}

This work was supported by the grants No. 21AG-1C047 and 24FP-3B021 of the
Higher Education and Science Committee of the Ministry of Education,
Science, Culture and Sport RA.

\appendix

\section{Evaluation of the integral over the momentum}

\label{sec:Int}

In this section we describe the transformation of the integral (\ref{Ixx})
over the momentum of the electromagnetic modes. In the first step of the
evaluation, we write the integral in the form%
\begin{equation}
I(x,x^{\prime })=\frac{1}{2}\int \frac{d\mathbf{k}}{\omega }\mathbf{\,}e^{i%
\mathbf{k}\cdot \Delta \mathbf{x}_{+}-i\omega \Delta t},  \label{Ixx2}
\end{equation}%
where $\Delta \mathbf{x}_{+}=(x_{+}^{1},\Delta \mathbf{x}_{\parallel })$, $%
\mathbf{k}$ $=(k^{1},\mathbf{k}_{\parallel })$, and the integrations go over 
$k^{i}\in (-\infty ,+\infty )$ for $i=1,\ldots ,D-1$. Introducing spherical
coordinates in the momentum space with $\mathbf{k}\cdot \Delta \mathbf{x}%
_{+}=k|\Delta \mathbf{x}_{+}|\cos \theta $, the integral over angular
coordinates is expressed in terms of the Bessel function:%
\begin{equation}
I(x,x^{\prime })=\frac{\left( 2\pi \right) ^{\frac{D-1}{2}}}{2|\Delta 
\mathbf{x}_{+}|^{\frac{D-3}{2}}}\int_{0}^{\infty }dk\mathbf{\,}k^{\frac{D-1}{%
2}}J_{\frac{D-3}{2}}\left( k|\Delta \mathbf{x}_{+}|\right) \frac{e^{-i\omega
\Delta t}}{\omega }.  \label{Ixx3}
\end{equation}%
Here, the integral is understood in the sense $\Delta t\rightarrow \Delta
t-i\epsilon /2$. The integral in (\ref{Ixx3}) is evaluated by using the
formula from \cite{Prud2} and we get 
\begin{equation}
I(x,x^{\prime })=\left( 2\pi \right) ^{\frac{D}{2}-1}\lambda ^{D-2}f_{\frac{D%
}{2}-1}\left( \lambda \sqrt{\left( \Delta \mathbf{x}_{+}\right) ^{2}-\left(
\Delta t\right) ^{2}+i\epsilon \Delta t}\right) ,  \label{Ixx4}
\end{equation}%
with the notation (\ref{fnuK}) for the Macdonald function. Note that in (\ref%
{Ixx4}), 
\begin{equation}
K_{\frac{D}{2}-1}(\lambda \sqrt{\left( \Delta \mathbf{x}_{+}\right)
^{2}-\left( \Delta t\right) ^{2}})=\frac{\pi }{2}e^{-\frac{\pi i}{4}D}H_{%
\frac{D}{2}-1}^{(2)}(\lambda \sqrt{\left( \Delta t\right) ^{2}-\left( \Delta 
\mathbf{x}_{+}\right) ^{2}}),  \label{H2}
\end{equation}%
for $|\Delta \mathbf{x}_{+}|<\Delta t$, and 
\begin{equation}
K_{\frac{D}{2}-1}(\lambda \sqrt{\left( \Delta \mathbf{x}_{+}\right)
^{2}-\left( \Delta t\right) ^{2}})=\frac{\pi }{2}e^{\frac{\pi i}{4}D}H_{%
\frac{D}{2}-1}^{(1)}(\lambda \sqrt{\left( \Delta t\right) ^{2}-\left( \Delta 
\mathbf{x}_{+}\right) ^{2}}).  \label{H1}
\end{equation}%
in the case $|\Delta \mathbf{x}_{+}|<|\Delta t|$, $\Delta t<0$. Here, $%
H_{\nu }^{(1,2)}(u)$ are the Hankel functions.

The Wightman function for a scalar field of mass $\lambda $ in $D$%
-dimensional Minkowski spacetime with Cartesian coordinates $%
x^{i}=(t,x^{1},\ldots ,x^{D-1})$ is given by%
\begin{equation}
W_{D}^{+}(\lambda ;x,x^{\prime })=\frac{\lambda ^{D-2}}{\left( 2\pi \right)
^{\frac{D}{2}}}f_{\frac{D}{2}-1}\left( \lambda \sqrt{-\eta _{il}\Delta
x^{i}\Delta x^{l}+i\epsilon \Delta t}\right) ,  \label{WD}
\end{equation}%
where $\Delta x^{i}=x^{i}-x^{\prime i}$ and $i,l=0,1,\ldots ,D-1$. Now we
see that the function $I(x,x^{\prime })$ is expressed in terms of the
function (\ref{WD}) as 
\begin{equation}
I(x,x^{\prime })=\left( 2\pi \right) ^{D-1}W_{D}^{+}(\lambda ;x,x_{\mathrm{im%
}-}^{i}),  \label{Ixx5}
\end{equation}%
where $x_{\mathrm{im}-}^{\prime }=(t^{\prime },-x^{\prime 1},x^{\prime
2},\ldots ,x^{\prime D-1})$ is the image of the point $x^{\prime }$ with
respect to the boundary $x^{1}=0$.


\begin{thebibliography}{99}
\bibitem{Hawk73} S.W. Hawking, G.F.R. Ellis, The Large Scale Structure of
Space-Time, Cambridge, Cambridge University Press, 1973.

\bibitem{Grif09} J.B. Griffiths, J. Podolsky, Exact Space-Times in
Einstein's General Relativity, Cambridge, Cambridge University Press, 2009.

\bibitem{Ahar00} O. Aharony, S.S. Gubser, J. Maldacena, H. Ooguri, Y. Oz,
Large N field theories, string theory and gravity, Phys. Rep. 323 (2000) 183.

\bibitem{Nast15} H. N\u{a}stase, Introduction to AdS/CFT Correspondence,
Cambridge, Cambridge University Press, 2015.

\bibitem{Ammo15} M. Ammon, J. Erdmenger, Gauge/Gravity Duality: Foundations
and Applications, Cambridge, Cambridge University Press, 2015.

\bibitem{Zaa15} J. Zaanen, Y.-W. Sun, Y. Liu, K. Schalm, Holographic Duality
in Condensed Matter Physics, Cambridge, Cambridge University Press, 2015.

\bibitem{Brax04} P. Brax, C. van de Bruck, A.-C. Davis, Brane world
cosmology, Rep. Prog. Phys. 67 (2004) 2183.

\bibitem{Maar10} R. Maartens, K. Koyama, Brane-world gravity, Living Rev.
Relativity 13 (2010) 5.

\bibitem{Most97} V. M. Mostepanenko, N.N. Trunov, The Casimir Effect and Its
Applications, Oxford, Clarendon, 1997.

\bibitem{Milt02} K.A. Milton, The Casimir Effect: Physical Manifestation of
Zero-Point Energy, Singapore, World Scientific, 2002.

\bibitem{Bord09} M. Bordag, G.L. Klimchitskaya, U. Mohideen, V.M.
Mostepanenko, Advances in the Casimir Effect, New York, Oxford University
Press, 2009.

\bibitem{Casi11} \textit{Casimir Physics}, edited by D. Dalvit, P. Milonni,
D. Roberts, F. da Rosa, Lecture Notes in Physics Vol. 834, Berlin,
Springer-Verlag, 2011.

\bibitem{Fabi00} M. Fabinger, P. Horava, Casimir effect between world-branes
in heterotic M-theory, Nucl. Phys. B 580 (2000) 243.

\bibitem{Noji00} S. Nojiri, S. Odintsov, S. Zerbini, Quantum (in)stability
of dilatonic AdS backgrounds and the holographic renormalization group with
gravity, Phys. Rev. D 62 (2000) 064006.

\bibitem{Toms00} D.J. Toms, Quantised bulk fields in the Randall--Sundrum
compactification model, Phys. Lett. B 484 (2000) 149.

\bibitem{Noji00b} S. Nojiri, O. Obregon, S. Odintsov, (Non)-singular
brane-world cosmology induced by quantum effects in five-dimensional
dilatonic gravity, Phys. Rev. D 62 (2000) 104003.

\bibitem{Gold00} W. D. Goldberger, I. Z. Rothstein, Quantum stabilization of
compactified AdS$_{5}$, Phys. Lett. B 491 (2000) 339.

\bibitem{Noji00c} S. Nojiri, S. Odintsov, Brane-world cosmology in higher
derivative gravity or warped compactification in the next-to-leading order
of AdS/CFT correspondence, J. High Energy Phys. 07 (2000) 049.

\bibitem{Garr01} J. Garriga, O. Pujol\'{a}s, T. Tanaka, Radion effective
potential in the brane-world, Nucl. Phys. B 605 (2001) 192.

\bibitem{Brev01} I.H. Brevik, K.A. Milton, S. Nojiri, S.D. Odintsov, Quantum
(in)stability of a brane-world AdS$_{5}$ universe at nonzero temperature,
Nucl. Phys. B 599 (2001) 305.

\bibitem{Flac01} A. Flachi, D.J. Toms, Quantized bulk scalar fields in the
Randall-Sundrum brane model, Nucl. Phys. B 610 (2001) 144.

\bibitem{Flac01b} A. Flachi, I.G. Moss, D.J. Toms, Fermion vacuum energies
in brane world models, Phys. Lett. B 518 (2001) 153.

\bibitem{Flac01c} A. Flachi, I.G. Moss, D.J. Toms, Quantized bulk fermions
in the Randall-Sundrum brane model, Phys. Rev. D 64 (2001) 105029.

\bibitem{Saha03} A.A. Saharian, M.R. Setare, The Casimir effect on
background of conformally flat brane-world geometries, Phys. Lett. B 552
(2003) 119.

\bibitem{Eliz03} E. Elizalde, S. Nojiri, S.D. Odintsov, S. Ogushi, Casimir
effect in de Sitter and anti-de Sitter braneworlds, Phys. Rev. D 67 (2003)
063515.

\bibitem{Flac03} A. Flachi, J. Garriga, O. Pujol\'{a}s, T. Tanaka, Moduli
stabilization in higher dimensional brane models, J. High Energy Phys. 08
(2003) 053.

\bibitem{Flac03b} A. Flachi, O. Pujol\'{a}s, Quantum self-consistency of AdS$%
\times \Sigma $ brane models Phys. Rev. D 68, 025023 (2003).

\bibitem{Garr03} J. Garriga, A. Pomarol, A stable hierarchy from Casimir
forces and the holographic interpretation Phys. Lett. B 560 (2003) 91.

\bibitem{Durr07} R. Durrer, M. Ruser, Dynamical Casimir effect in
braneworlds, Phys. Rev. Lett. 99 (2007) 071601.

\bibitem{Ruse07} M. Ruser, R. Durrer, Dynamical Casimir effect for gravitons
in bouncing braneworlds, Phys. Rev. D 76 (2007) 104014.

\bibitem{Eliz07} E. Elizalde, M. Minamitsuji, W. Naylor, Casimir effect in
rugby-ball type flux compactifications, Phys. Rev. D 75 (2007) 064032.

\bibitem{Saha07} A.A. Saharian, A.L. Mkhitaryan, Wightman function and
vacuum densities for a Z$_{2}$-symmetric thick brane in AdS spacetime, J.
High Energy Phys. 08 (2007) 063.

\bibitem{Fran07} M. Frank, I. Turan, L. Ziegler, Casimir force in
Randall-Sundrum models, Phys. Rev. D 76 (2007) 015008.

\bibitem{Lina08} R. Linares, H.A. Morales-T\'{e}cotl, O. Pedraza, Casimir
force for a scalar field in warped brane worlds, Phys. Rev. D 77 (2008)
066012.

\bibitem{Fran08} M. Frank, N. Saad, I. Turan, Casimir force in
Randall-Sundrum models with d+1 dimensions, Phys. Rev. D 78 (2008) 055014.

\bibitem{Flac09} A. Flachi, T. Tanaka, Casimir effect on the brane, Phys.
Rev. D 80 (2009) 124022.

\bibitem{Teo09} L.P. Teo, Casimir effect in spacetime with extra dimensions
- from Kaluza-Klein to Randall-Sundrum models, Phys. Lett. B 682 (2009) 259.

\bibitem{Rype10} M. Rypestol, I. Brevik, Finite-temperature Casimir effect
in Randall-Sundrum models, New J. Phys. 12 (2010) 013022.

\bibitem{Obous11} R. Obousy, G. Cleaver, Casimir energy and brane stability,
J. Geom. Phys. 61 (2011) 577.

\bibitem{Teo13} L.P. Teo, Finite temperature fermionic Casimir interaction
in anti-de Sitter space-time, Int. J. Mod. Phys. A 28 (2013) 1350158.

\bibitem{Knap04} R.A. Knapman, D.J. Toms, Stress-energy tensor for a
quantized bulk scalar field in the Randall-Sundrum brane model, Phys. Rev. D
69 (2004) 044023.

\bibitem{Saha05} A.A. Saharian, Wightman function and Casimir densities on
AdS bulk with application to the Randall-Sundrum braneworld, Nucl. Phys. B
712 (2005) 196.

\bibitem{Saha06} A.A. Saharian, Wightman function and vacuum fluctuations in
higher dimensional brane models, Phys. Rev. D 73 (2006) 044012.

\bibitem{Saha06b} A.A. Saharian, Bulk Casimir densities and vacuum
interaction forces in higher dimensional brane models, Phys. Rev. D 73
(2006) 064019.

\bibitem{Shao10} S.-H. Shao, P. Chen, J.-A. Gu, Stress-energy tensor induced
by a bulk Dirac spinor in the Randall-Sundrum model, Phys. Rev. D 81 (2010)
084036.

\bibitem{Eliz13} E. Elizalde, S.D. Odintsov, A.A. Saharian, Fermionic
Casimir densities in anti-de Sitter spacetime, Phys. Rev. D 87 (2013) 084003.

\bibitem{Bell22AdS} S. Bellucci, W. Oliveira dos Santos, E.R. Bezerra de
Mello, A.A. Saharian, Cosmic string and brane induced effects on the
fermionic vacuum in AdS spacetime, J. High Energy Phys. 05 (2022) 021.

\bibitem{Saha04} A.A. Saharian, Surface Casimir densities and induced
cosmological constant on parallel branes in AdS spacetime, Phys. Rev. D 70
(2004) 064026.

\bibitem{Saha06c} A.A. Saharian, Surface Casimir densities and induced
cosmological constant in higher dimensional braneworlds, Phys. Rev. D 74
(2006) 124009.

\bibitem{Teo10} L.P. Teo, Casimir effect of electromagnetic field in
Randall-Sundrum spacetime, J. High Energy Phys. 10 (2010) 019.

\bibitem{Kota17} A.S. Kotanjyan, A.A. Saharian, Electromagnetic quantum
effects in anti-de Sitter spacetime, Phys. At. Nucl. 80 (2017) 562.

\bibitem{Saha20b} A.A. Saharian, A.S. Kotanjyan, A.A. Saharyan, H. G.
Sargsyan, Electromagnetic Casimir densities for planar boundaries in AdS
spacetime, Int. J. Mod. Phys. A 35 (2020) 2040029.

\bibitem{Saha20} A.A. Saharian, A.S. Kotanjyan, H.G. Sargsyan,
Electromagnetic field correlators and the Casimir effect for planar
boundaries in AdS spacetime with application in braneworlds, Phys. Rev. D
102 (2020) 105014.

\bibitem{Taka11} T. Takayanagi, Holographic dual of BCFT, Phys. Rev. Lett.
107 (2011) 101602.

\bibitem{Fuji11} M. Fujita, T. Takayanagi, E. Tonni, Aspects of AdS/BCFT, J.
High Energy Phys. 11 (2011) 043.

\bibitem{Karc01} A. Karch, L. Randall, Open and closed string interpretation
of SUSY CFT's on branes with boundaries, J. High Energy Phys. 06 (2001) 063.

\bibitem{Suzu24} K. Suzuki, Gauge symmetries and conserved currents in
AdS/BCFT, J. High Energy Phys. 06 (2024) 137.

\bibitem{Fu25} Y. Fu, K.-Y. Kim, Wedge holographic complexity in
Karch-Randall braneworld, J. High Energy Phys. 01 (2025) 174.

\bibitem{Ahn25} B. Ahn, S.-E. Bak, K-Y. Kim, M. Nishida, Renyi reflected
entropy and entanglement wedge cross section with cosmic branes in AdS/BCFT,
Phys. Rev. D 111 (2025) 126006.

\bibitem{Hao25} P.-X. Hao, N. Ogawa, T. Takayanagi, T. Waki, Flat space
solography via AdS/BCFT, arXiv:2509.00652.

\bibitem{Ryu06} S. Ryu, T. Takayanagi, Holographic derivation of
entanglement entropy from the anti-de Sitter space/Conformal field theory
correspondence, Phys. Rev. Lett. 96 (2006) 181602.

\bibitem{Ryu06b} S. Ryu, T. Takayanagi, Aspects of holographic entanglement
entropy, J. High Energy Phys. 08 (2006) 045.

\bibitem{Nish09} T. Nishioka, S. Ryu, T. Takayanagi, Holographic
entanglement entropy: an overview, J. Phys. A 42 (2009) 504008.

\bibitem{Chen22} B. Chen, B. Czech, Zi-Zhi Wang, Quantum information in
holographic duality, Rep. Prog. Phys. 85 (2022) 046001.

\bibitem{Beze15} E.R. Bezerra de Mello, A.A. Saharian, M.R. Setare, Vacuum
densities for a brane intersecting the AdS boundary, Phys. Rev. D 92 (2015)
104005.

\bibitem{Bell22} S. Bellucci, A.A. Saharian, V.Kh. Kotanjyan, Vacuum
densities and the Casimir forces for branes orthogonal to the AdS boundary,
Phys. Rev. D 106 (2022) 065021.

\bibitem{Saha23} A.A. Saharian, Surface Casimir densities on branes
orthogonal to the boundary of anti-de Sitter spacetime, Physics 5(4) (2023)
1145.

\bibitem{Davi81} P.C.W. Davies, S.D. Unwin, Quantum vacuum energy and the
masslessness of the photon, Phys. Lett. B 98 (1981) 274.

\bibitem{Bart85} G. Barton, The Casimir effect with finite mass photons,
Ann. Phys. 162 (1985) 231.

\bibitem{Saha26} A.A. Saharian, H.H. Asatryan, Two-point functions and the
vacuum densities in the Casimir effect for the Proca field, Eur. Phys. J.
Plus 141 (2026) 106.

\bibitem{Alle86} B. Allen,T. Jacobson, Vector two point functions in
maximally symmetric spaces, Commun. Math. Phys. 103 (1986) 669.

\bibitem{Liu99} H. Liu, A.A. Tseytlin, On four-point functions in the
CFT/AdS correspondence, Phys. Rev. D 59 (1999) 086002.

\bibitem{Hoke99} E. D'Hoker, D. Z. Freedman, Gauge boson exchange in AdS$%
_{d+1}$, Nucl. Phys. B 544 (1999) 612.

\bibitem{Prud2} A.P. Prudnikov, Yu.A. Brychkov, O.I. Marichev, Integrals and
Series, New York, Gordon and Breach, 1986, Vol. 2.

\bibitem{Alis11} M. Alishahiha, R. Fareghbal, Boundary CFT from holography,
Phys. Rev. D 84 (2011) 106002.

\bibitem{Hint15} K. Hinterbichler, J. Stokes, M. Trodden, Holographic CFTs
on maximally symmetric spaces: Correlators, integral transforms and
applications, Phys. Rev. D 92 (2015) 065025.

\bibitem{Birr82} N.D. Birrell, P.C.W. Davies, Quantum Fields in Curved
Space, Cambridge, Cambridge University Press, 1982.

\bibitem{Park09} L. Parker, D. Toms, Quantum Field Theory in Curved
Spacetime, Cambridge, Cambridge University Press, 2009.
\end{thebibliography}
\end{document}